\documentclass[journal]{IEEEtran}

\ifCLASSINFOpdf
  \usepackage[pdftex]{graphicx}
  \DeclareGraphicsExtensions{.pdf,.jpeg,.png}
\else
  \usepackage[dvips]{graphicx}
\fi
\usepackage{subfigure}
\usepackage{epstopdf}
\usepackage[cmex10]{amsmath}
\usepackage{amssymb}
\usepackage{wasysym}
\usepackage{algorithm}
\usepackage{algorithmic}
\usepackage{array}
\usepackage{cite}
\usepackage{color}
\usepackage{url}
\usepackage{booktabs}

\usepackage{epsfig,latexsym}
\usepackage{flushend}
\usepackage{cite}

\begin{document}
%
\title{Beam Training and Allocation for Multiuser Millimeter Wave Massive MIMO Systems}

\author{Xuyao~Sun,~\IEEEmembership{Student~Member,~IEEE}, Chenhao~Qi,~\IEEEmembership{Senior~Member,~IEEE}, \\ and Geoffrey Ye Li,~\IEEEmembership{Fellow,~IEEE}
\thanks{This work is supported in part by National Natural Science Foundation of China under Grant 61871119 and Natural Science Foundation of Jiangsu Province under Grant BK20161428. (\textit{Corresponding author: Chenhao~Qi})}
\thanks{Xuyao~Sun and Chenhao~Qi are with the School of Information Science and Engineering, Southeast University, Nanjing 210096, China (Email: qch@seu.edu.cn).}
\thanks{Geoffrey Ye Li is with the School of Electrical and Computer Engineering, Georgia Institute of Technology, Atlanta, GA, USA (Email: liye@ece.gatech.edu).}
}
\markboth{}
{Shell \MakeLowercase{\textit{et al.}}: Bare Demo of IEEEtran.cls for Journals}
\maketitle
\vspace{-50 pt}
\begin{abstract}
We investigate beam training and allocation for multiuser millimeter wave massive MIMO systems. An orthogonal pilot based beam training scheme is first developed to reduce the number of training times, where all users can simultaneously perform the beam training with the base station (BS). As the number of users increases, the same beam from the BS may point to different users, leading to beam conflict and multiuser interference. Therefore, a quality-of-service (QoS) constrained (QC) beam allocation scheme is proposed to maximize the equivalent channel gain of the QoS-satisfied users, under the premise that the number of the QoS-satisfied users without beam conflict is maximized. To reduce the overhead of beam training, two partial beam training schemes, an interlaced scanning (IS) and a selection probability (SP) based schemes, are proposed. The overhead of beam training for the IS-based scheme can be reduced by nearly half while the overhead for the SP-based scheme is flexible. Simulation results show that the QC-based beam allocation scheme can effectively mitigate the interference caused by the beam conflict and significantly improve the spectral efficiency while the IS-based and SP-based schemes significantly reduce the overhead of beam training at the cost of sacrificing spectral efficiency a little.
\end{abstract}

\section{Introduction}
Millimeter wave (mmWave) communications, ranging from 30GHz to 300GHz frequency band, has been regarded as a promising technology for future wireless systems~\cite{heath2016overview,wang2015multi,wei2014key} since it can considerably increase the data rate owing to its wider bandwidth. However, mmWave communication faces the challenge of high path loss caused by high carrier frequency~\cite{rappaport2013millimeter}. Fortunately, large antenna arrays can be packed into small form factors at mmWave frequencies~\cite{hong2014study}, making it feasible for both base station (BS) and user equipment to compensate for the high path loss~\cite{rusek2013scaling}.
In the conventional MIMO systems, a fully digital baseband precoding is usually used. However, in mmWave massive MIMO systems, fully digital precoding is impractical since the number of antennas is large and the working frequency is much higher than that of conventional MIMO systems~\cite{el2014spatially}. In this context, hybrid precoding, including analog and digital precoding, is usually adopted for mmWave massive MIMO communications~\cite{Wang2018Spatial,Yu2017Hybrid,Zhao2017Multi,zhao2017Tone}. Analog precoding, also known as analog beamforming, can produce directional transmission under the constraints of constant amplitude (CA) and limited resolution from phase shifters~\cite{wang2009beam,Kim2014Fast,Lin2017ANew,sun2017Weighted}. Digital precoding is used to multiplex independent data streams and to mitigate interference, which is similar to that in the fully digital sub-6GHz wireless systems.

There have been many hybrid precoding schemes and algorithms for single-user multi-stream mmWave systems. In~\cite{alkhateeb2014channel}, a hybrid precoding algorithm has been proposed to exploit the sparse property of mmWave channels in the angle domain, where the sum-rate maximizing problem is formulated and solved as a sparse reconstruction problem. In~\cite{xiao2016hierarchical}, \cite{xiao2017channel}, a hierarchical codebook design and a multi-beam search scheme for single-user multi-stream communications have been developed to acquire multiple beams quickly, where each beam is essentially formed by analog precoding. In~\cite{yu2016alternating}, the hybrid precoding design is treated as a matrix factorization problem and three alternating minimization algorithms have been proposed for fully-connected and partially-connected hybrid precoding structures.

In multiuser multi-stream mmWave systems, the BS simultaneously serves multiple users, each equipped with an mmWave antenna array. In~\cite{alkhateeb2015limited}, a low-complexity multiuser hybrid algorithm has been proposed for the precoder at the BS and the combiners at the user equipment with a small number of training and feedback. There may be severe performance degradation when the spatially multiplexed users are close in angle domain. To address the issue, only a small subset of users are served simultaneously to reduce the multiuser interference~\cite{yuan2016novel}. In~\cite{gao2016near}, a near-optimal (NO) beam selection algorithm can mitigate multiuser interference for mmWave massive MIMO systems, where users are classified into two groups: interference-users (IUs) and non-interference-users (NIUs). The beams with full power serve the NIUs while the appropriate beams are selected to serve the IUs with proper power to maximize the sum-rate. In~\cite{amadori2015low}, three beam selection algorithms with low RF-complexity have been developed for beamspace mmWave massive MIMO systems, where each user served by the BS is equipped with a single omnidirectional antenna. In~\cite{gao2016near,amadori2015low}, beam allocation is for beamspace mmWave massive MIMO systems equipped with discrete lens arrays (DLA). In fact, beam allocation can be also used for mmWave massive MIMO systems with uniform linear arrays (ULA) even if there is only limit literature on the topic.

In this paper, we consider the beam training and allocation for multiuser multi-stream mmWave massive MIMO systems. We first propose an orthogonal pilot (OP) based beam training scheme where all users can simultaneously perform the beam training. As the number of users increases, the same beam from the BS may point to different users, leading to beam conflict and multiuser interference. Therefore, we develop a quality-of-service (QoS) constrained beam allocation scheme, which aims at maximizing the equivalent channel gain of the QoS-satisfied users, under the premise that the number of the QoS-satisfied users without any beam conflict is maximized. To substantially reduce the overhead of beam training, we propose two partial beam training schemes, an interlaced scanning (IS) and a selection probability (SP) based beam training schemes. The overhead of beam training for the IS-based scheme can be reduced by nearly half while that for the SP-based scheme is flexible and can be set arbitrarily.

The rest of the paper is organized as follows. The problem is formulated in Section~\ref{sec.ProblemFormulation}. In Section~\ref{sec.HybridPrecoding}, the hybrid precoder design is investigated and the OP-based beam training scheme is proposed. In Section~\ref{sec.BeamAllocation}, the QoS constrained (QC) beam allocation scheme is developed. In Section~\ref{sec.BeamTrainingScheme}, two partial beam training schemes, including the IS-based and SP-based schemes, are introduced. Simulation results are provided in Section~\ref{sec.SimulationResults}. Finally Section~\ref{sec.conclusion} concludes the paper.

The notations used in this paper are as follows. Symbols for matrices (upper case) and vectors (lower case) are
in boldface. According to the convention, $a$, $\boldsymbol{a}$, $\boldsymbol{A}$, and $\boldsymbol{\mathcal{A}}$ denote a scalar, a vector, a matrix, and a set, respectively. $[\boldsymbol{a}]_{i}$, $[\boldsymbol{A}]_{i,:}$, $[\boldsymbol{A}]_{:,j}$, and $[\boldsymbol{A}]_{i,j}$ represent the $i$th entry of $\boldsymbol{a}$, the $i$th-row of $\boldsymbol{A}$, the $j$th-column of $\boldsymbol{A}$, and the entry on the $i$th-row and $j$th-column of $\boldsymbol{A}$, respectively. $(\cdot)^T $, $(\cdot)^* $, $(\cdot)^H $, $(\cdot)^{-1}$, $\|\cdot\|_0$, and $\|\cdot\|_2$ denote the transpose, the conjugate, the conjugate transpose (Hermitian), the inverse, the zero norm, and the two norm (or Euclidean norm), respectively. $\boldsymbol{0}_{K}$, $\boldsymbol{I}_{K}$, and $\varnothing$ are the zero vector of size $K$, the identity matrix of size $K$, and the empty set, respectively. $\mathcal{CN}(m,\boldsymbol{R})$ is the complex Gaussian distribution with the mean of $m$ and the covariance matrix $\boldsymbol{R}$. ${\mathbb{E}}[\cdot]$ denotes the expectation. $\mathbb{C}$ is the set of complex number.



\section{Problem Formulation}\label{sec.ProblemFormulation}
\begin{figure}[!t]
\centering
\includegraphics[width=85mm,height=40mm]{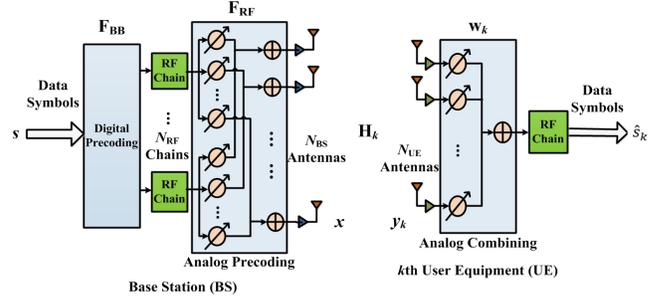}
\caption{A BS with hybrid precoding communicating with the $k$th user that employs analog-only combining in the downlink.}
\label{fig:hybrid}
\end{figure}

As shown in Fig.~\ref{fig:hybrid}, consider a multiuser mmWave massive MIMO communication system with a BS and $K$ users. The BS is with $N_{\rm BS}$ ULA antennas and $N_{\rm RF}$ RF chains ($N_{\rm BS}\gg N_{\rm RF} \geqslant 1$) and employs hybrid precoding while each user equipment (UE) is with $N_{\rm UE}$ ULA antennas and a single RF chain and employs an analog-only combining architecture. The maximum number of users simultaneous served by the BS is restricted by the number of its RF chains, i.e., $K\leq N_{\rm RF}$. It is commonly assumed that the number of simultaneously served users is the same as that of RF chains to save power consumption. If $K < N_{\rm RF}$, the BS will use $K$ out of $N_{\rm RF}$ RF chains to serve the $K$ users while turning off $(N_{\rm RF} -K)$ RF chains to save the BS power.

During the downlink transmission, the signal received by the $k$th user is denoted by
\begin{equation}\label{receivedvector}
\boldsymbol{y}_{k}=\boldsymbol{H}^{\rm dl}_{k}\boldsymbol{x}+\boldsymbol{n}^{\rm dl}_{k},
\end{equation}
where $\boldsymbol{x}=\boldsymbol{F}_{\rm RF}\boldsymbol{F}_{\rm BB}\boldsymbol{s}$ denotes the transmitted signal from the BS. $\boldsymbol{s} \triangleq{[s_1,s_2,\ldots,s_K]}^{T}$ is the data symbol vector subject to the constraint of total transmit power $P_{\rm dl}$, i.e., $\mathbb{E}[\boldsymbol{s}\boldsymbol{s}^{H}]=\frac{P_{\rm dl}}{K}\boldsymbol{I}_{K}$. $\boldsymbol{F}_{\rm BB} \triangleq [\boldsymbol{f}^{\rm BB}_{1},\boldsymbol{f}^{\rm BB}_{2},\ldots,\boldsymbol{f}^{\rm BB}_{K}] \in \mathbb{C}^{K \times K } $ and $\boldsymbol{F}_{\rm RF} \triangleq [\boldsymbol{f}^{\rm RF}_{1},\boldsymbol{f}^{\rm RF}_{2},\ldots,\boldsymbol{f}^{\rm RF}_{K}] \in \mathbb{C}^{N_{\rm BS} \times K} $ are the baseband precoder (digital precoder) and RF precoder (analog precoder), respectively. $\boldsymbol{H}^{\rm dl}_{k} \in \mathbb{C}^{N_{\rm UE}\times N_{\rm BS}}$ is the channel matrix between the BS and the $k$th user. $\boldsymbol{n}^{\rm dl}_{k}\sim\mathcal{CN}(0,\sigma^2_{\rm dl}\boldsymbol{I}_{N_{\rm UE}})$ denotes the noise term with each entry obeying independent complex Gaussian distribution with zero mean and variance of $\sigma^2_{\rm dl}$.

After the RF combiner $\boldsymbol{w}_{k}$ at the $k$th user, we have
\begin{equation}\label{finaldata2}
\hat{s}_{k} =\boldsymbol{w}^{H}_{k}\boldsymbol{H}^{\rm dl}_{k}\boldsymbol{x}+\boldsymbol{w}^{H}_{k}\boldsymbol{n}^{\rm dl}_{k}=\boldsymbol{w}^{H}_{k}\boldsymbol{H}^{\rm dl}_{k}\boldsymbol{F}_{\rm RF}\sum_{n=1}^{K}\boldsymbol{f}^{\rm BB}_{n}s_{n}+\boldsymbol{w}^{H}_{k}\boldsymbol{n}^{\rm dl}_{k}.
\end{equation}

Note that $\boldsymbol{F}_{\rm RF}$ and $\boldsymbol{w}_{k}$ in Fig.~\ref{fig:hybrid} are implemented using phase shifters. The entries of $\boldsymbol{F}_{\rm RF}$ and $\boldsymbol{w}_{k}$ have constant envelop. Furthermore, the angles of the phase shifters usually have a finite set of possible values. With these constraints, we have $[\boldsymbol{F}_{\rm RF}]_{m,n}=\frac{1}{\sqrt{N_{\rm BS}}}e^{j\phi_{m,n}}$, $[\boldsymbol{w}_{k}]_{m}=\frac{1}{\sqrt{N_{\rm UE}}}e^{j\theta_m}$, where $\phi_{m,n}$ and $\theta_m$ are the quantized angles. Moreover, we normalize $\boldsymbol{F}_{\rm BB}$ such that $\|\boldsymbol{F}_{\rm RF}\boldsymbol{F}_{\rm BB}\|^2_F=K$, that is, the baseband precoder provides no power gain.

According to the existing literature~\cite{rappaport2013millimeter,rappaport2013broadband}, mmWave channels have limited scatterers. We adopt a widely used geometric channel model with $L_k$ scatterers. The ULA with a half-wavelength antenna space is equipped at the BS and the users. Then the $N_{\rm UE}\times N_{\rm BS}$ channel matrix between the BS and the $k$th user can be expressed as
\begin{equation}\label{channelmodel}
\boldsymbol{H}^{\rm dl}_k=\sqrt{\frac{N_{\rm BS}N_{\rm UE}}{L_k}}\sum_{l=1}^{L_k}\beta^k_l\boldsymbol{a}_{\rm UE}(\theta^k_l)\boldsymbol{a}^{H}_{\rm BS}(\phi^k_l)
\end{equation}
where $\beta^k_l$ is the complex gain of the $l$th path with ${\mathbb{E}}[|\beta^k_l|^2]=\bar\beta$. $\theta^k_l \triangleq {\rm sin}(\vartheta^k_l)$ and $\phi^k_l \triangleq {\rm sin}(\varphi^k_l)$ are the angle-of-arrival (AoA) and the angle-of-departure (AoD) of the $l$th path, respectively. $\vartheta^k_l, \varphi^k_l \in [-\pi/2, \pi/2]$ are the physical angles for the AoA and the AoD, respectively. The channel steering vectors at the BS and the $k$th user are denoted as $\boldsymbol{a}_{\rm BS}(\phi^k_l)=\boldsymbol{u}(N_{\rm BS},\phi^k_l)$ and $\boldsymbol{a}_{\rm UE}(\theta^k_l)=\boldsymbol{u}(N_{\rm UE},\theta^k_l)$, respectively. $\boldsymbol{u}(N,\alpha)$ is defined as
\begin{equation}\label{steeringvector}
\boldsymbol{u}(N,\alpha)\triangleq\frac{1}{\sqrt{N}}{[1,e^{j\pi\alpha},\ldots,e^{j(N-1)\pi\alpha}]}^T.
\end{equation}

Our objective is to design the hybrid precoder (analog and digital precoders) at the BS and the analog combiner at each user to maximize the sum-rate of the system, a commonly adopted performance metric. Based on (\ref{finaldata2}), we can write the achievable rate of the $k$th user as
\begin{equation}\label{AchievableRate}
R_k=\log_{2}\Bigg(1+\frac{\frac{P}{K}|\boldsymbol{w}^H_k\boldsymbol{H}^{\rm dl}_k\boldsymbol{F}_{\rm RF}\boldsymbol{f}^{\rm BB}_k|^2}{\frac{P}{K}\sum_{i\neq k}|\boldsymbol{w}^H_k\boldsymbol{H}^{\rm dl}_k\boldsymbol{F}_{\rm RF}\boldsymbol{f}^{\rm BB}_i|^2+\sigma^2_{\rm dl}}\Bigg).
\end{equation}
The sum-rate of the system is $R_{\rm sum}=\sum_{k=1}^KR_k$.

The analog precoder is formed by some codewords selected from a beam steering codebook, which essentially consists of $N_{\rm BS}$ equally spaced channel steering vectors pointing at $N_{\rm BS}$ different directions~\cite{ShiwenTSP2017}. Similarly, the analog combiner is formed by some codewords selected from a beam steering codebook which consists of $N_{\rm UE}$ channel steering vectors. The codebooks at the BS and the users are denoted by $\boldsymbol{\mathcal{F}}_c=\{\boldsymbol{f}_c(1),\boldsymbol{f}_c(2),...,\boldsymbol{f}_c(N_{\rm BS})\}$ and $\boldsymbol{\mathcal{W}}_c=\{\boldsymbol{w}_c(1),\boldsymbol{w}_c(2),...,\boldsymbol{w}_c(N_{\rm UE})\}$, respectively, where
\begin{align}\label{BSbeamformingvector}
\boldsymbol{f}_c(n) &= \boldsymbol{u}(N_{\rm BS},-1+(2n-1)/N_{\rm BS}), \nonumber \\
\boldsymbol{w}_c(n) &= \boldsymbol{u}(N_{\rm UE},-1+(2n-1)/N_{\rm UE}).
\end{align}

Then the sum-rate maximization problem in terms of $\boldsymbol{F}_{\rm RF}$, $\boldsymbol{F}_{\rm BB}$, and $\boldsymbol{w}_k$ can be formulated as
\begin{align}\label{OptimizationProblem}
&  \max_{\substack{ \boldsymbol{F}_{\rm RF}, \boldsymbol{F}_{\rm BB},\\ \boldsymbol{w}_k} } \sum_{k=1}^K\log_{2}\Bigg(1+\frac{\frac{P_{\rm dl}}{K}|\boldsymbol{w}^H_k\boldsymbol{H}^{\rm dl}_k\boldsymbol{F}_{\rm RF}\boldsymbol{f}^{\rm BB}_k|^2}{\frac{P_{\rm dl}}{K}\sum_{i\neq k}|\boldsymbol{w}^H_k\boldsymbol{H}^{\rm dl}_k\boldsymbol{F}_{\rm RF}\boldsymbol{f}^{\rm BB}_i|^2+\sigma^2_{\rm dl}}\Bigg)\notag\\
&~~~~{\rm s.t.}  \ \ \ [\boldsymbol{F}_{\rm RF}]_{:,k}=\boldsymbol{f}^{\rm RF}_k\in\boldsymbol{\mathcal{F}}_c,~k=1,2,...,K,\notag\\
& \ \ \ \ \ \ \ \ \ \ \boldsymbol{w}_k\in\boldsymbol{\mathcal{W}}_c,~k=1,2,...,K,\notag\\
& \ \ \ \ \ \ \ \ \ \ {\|\boldsymbol{F}_{\rm RF}\boldsymbol{f}^{\rm BB}_k\|}^2_F=1,~ k=1,2,...,K.
\end{align}
Note that the design of the analog precoder and digital precoder is coupling. Each computation of the
sum-rate can be performed only after both the analog precoder and digital precoder are determined, which leads to intractable computational complexity to solve (7). According to the existing literature~\cite{Zhao2017Multi,zhao2017Tone,xiao2017channel,alkhateeb2015limited,yuan2016novel}, a typical method to decouple the design of the analog precoder and digital precoder is first determining $\boldsymbol{F}_{\rm RF}$ and $\{\boldsymbol{w}_k\}^K_{k=1}$ while fixing $\boldsymbol{F}_{\rm BB}$, and then determining $\boldsymbol{F}_{\rm BB}$ based on zero-forcing (ZF) or minimum mean-squared error (MMSE) criterion. When determining $\boldsymbol{F}_{\rm RF}$ and $\{\boldsymbol{w}_k\}^K_{k=1}$, the optimization problem can be written as
\begin{align}\label{DetermineFirst}
& \max_{\boldsymbol{f}^{\rm RF}_k, \boldsymbol{w}_k} |\boldsymbol{w}^H_k\boldsymbol{H}^{\rm dl}_k\boldsymbol{f}_k^{\rm RF}| ,~~~k=1,2,...,K,\\
 &~~{\rm s.t.}~~ \boldsymbol{w}_k\in\boldsymbol{\mathcal{W}}_c,~\boldsymbol{f}^{\rm RF}_k\in\boldsymbol{\mathcal{F}}_c,\notag
\end{align}
where $|\boldsymbol{w}^H_k\boldsymbol{H}^{\rm dl}_k\boldsymbol{f}_k^{\rm RF}|$ is called as the equivalent channel gain. Beam training is based on exhaustive search to solve (\ref{DetermineFirst}) for the best analog precoder and the best analog combiner to maximize the equivalent channel gain for each user, which requires $N_{\rm BS } N_{\rm UE}K / N_{\rm RF}$ times of beam training, e.g., $N_{\rm BS } N_{\rm UE}K / N_{\rm RF}= 512$ if $N_{\rm BS }=64$, $N_{\rm UE}=8$, $N_{\rm RF}=8$ and  $K=8$.


\section{Hybrid Precoder Design}\label{sec.HybridPrecoding}
The hybrid precoder design is usually divided into two stages. The first stage includes the beam training and analog precoder design. The second stage contains the channel estimation and digital precoder design. In this section, we first propose an OP-based beam training scheme, which reduces the number of beam training from $N_{\rm BS } N_{\rm UE}K / N_{\rm RF}$ to $N_{\rm BS} N_{\rm UE} / N_{\rm RF}$. Then we present a channel estimation method for the OP-based beam training scheme.

\subsection{OP-based Beam Training and Analog Precoder Design}\label{subsec.BeamTraining}
For time-division duplex (TDD) systems, the channel reciprocity holds, i.e., $\boldsymbol{H}^{\rm dl}_k=(\boldsymbol{H}^{\rm ul}_k)^T$, where the superscript ``${\rm ul}$'' is short for uplink and $\boldsymbol{H}^{\rm ul}_k$ denotes the uplink $N_{\rm BS}\times N_{\rm UE}$ channel matrix for the $k$th user. In the OP-based beam training scheme, the users transmit mutually orthogonal pilot sequences so that the signals from different users can be distinguished at the BS. The pilot sequences are denoted as $\sqrt{\tau P_{\rm ul}}\boldsymbol{\phi}_k\in \mathbb{C}^{1\times \tau},k=1,2,...,K$, with $\boldsymbol{\phi}_k\boldsymbol{\phi}^H_j=\delta[k-j]$, where $\tau(\tau\geq K)$ is the length of the pilot sequence, $\delta[n]$ is a Dirac delta function, and $\delta[n]=1$ if $n=0$, $\delta[n]=0$ if $n\neq0$. $P_{\rm ul}$ is the uplink transmit power of each user.

Beam training will repeat $N_{\rm BS}N_{\rm UE}/N_{\rm RF}$ times for different combinations of codewords for the users and the BS. During beam training, all users select the same codeword from $\boldsymbol{\mathcal{W}}_c$, e.g., $\boldsymbol{w}_c(n)$ for $n=1,2,\ldots,N_{\rm UE}$, as the analog beamforming vector. Note that if all users select the same codeword from $\boldsymbol{\mathcal{W}}_c$, it is easier for the BS to record which codewords have been tested and which codewords have not yet been tested, where the record is the same for different users. Given $\boldsymbol{w}_c(n)$, the BS can select $N_{\rm RF}$ different codewords from $\boldsymbol{\mathcal{F}}_c$ instead of only one codeword each time since the BS has $N_{\rm RF}$ RF chains and repeat $N_{\rm BS}/N_{\rm RF}$ times. The selected $N_{\rm RF}$ codewords in the $m$th selection form a $N_{\rm BS}\times N_{\rm RF}$ analog combining matrix $\boldsymbol{F}_{\rm RF}^{(m)}$. Then the $N_{\rm RF}\times \tau$ received signal matrix at the BS in the $(n,m)$th beam training is a summation of the pilots from all users and can be expressed as
\begin{equation}\label{BSReceivedPilotSignal}
\boldsymbol{Y}^{(m,n)}=\sum_{k=1}^K\sqrt{\tau P_{\rm ul}}(\boldsymbol{F}^{(m)}_{\rm RF})^H\boldsymbol{H}^{\rm ul}_k\boldsymbol{w}_c(n)\boldsymbol{\phi}_k + (\boldsymbol{F}^{(m)}_{\rm RF})^H \boldsymbol{N}^{\rm ul},
\end{equation}
for $m=1,2,...,N_{\rm BS}/N_{\rm RF}$ and $n=1,2,...,N_{\rm UE}$. $\boldsymbol{N}^{\rm ul}$ represents the uplink channel noise, with each entry obeys the complex Gaussian distribution with zero-mean and variance of $\sigma^2_{\rm ul}$.

To get channel information corresponding to the $k$th user, multiply $\boldsymbol{Y}^{(m,n)}$ with the conjugate of $\boldsymbol{\phi}_k$ on the left and then
\begin{align}\label{ProcessedPilotSignal}
\boldsymbol{r}^{(m,n)}_k&=\frac{\boldsymbol{\phi}^*_k}{\sqrt{\tau P_{\rm ul}}}(\boldsymbol{Y}^{(m,n)})^T \notag \\
&=\boldsymbol{w}^T_c(n)(\boldsymbol{H}^{\rm ul}_k)^T(\boldsymbol{F}^{(m)}_{\rm RF})^*+\frac{\boldsymbol{\phi}^*_k(\boldsymbol{N}^{\rm ul})^T}{\sqrt{\tau P_{\rm ul}}} (\boldsymbol{F}^{(m)}_{\rm RF})^* \notag\\
&=\boldsymbol{w}^T_c(n)\boldsymbol{H}^{\rm dl}_k(\boldsymbol{F}^{(m)}_{\rm RF})^*+\frac{\boldsymbol{\phi}^*_k(\boldsymbol{N}^{\rm ul})^T}{\sqrt{\tau P_{\rm ul}}} (\boldsymbol{F}^{(m)}_{\rm RF})^*,
\end{align}
where we have used the identity $\boldsymbol{\phi}_k\boldsymbol{\phi}^H_j=\delta[k-j]$. It is obvious that $\boldsymbol{r}^{(m,n)}_k$ is a $1\times N_{\rm RF}$ vector. For the OP-based beam training scheme, we put together $\boldsymbol{r}^{(m,n)}_k, m=1,2,\ldots,N_{\rm BS}/N_{\rm RF}, n=1,2,\ldots,N_{\rm UE}$ and obtain an $N_{\rm UE}\times N_{\rm BS}$ matrix
\begin{equation}\label{AllProcessedPilotSignal}
\boldsymbol{R}_k=\left[
\begin{matrix}
\boldsymbol{r}^{(1,1)}_k & \boldsymbol{r}^{(2,1)}_k & \cdots & \boldsymbol{r}^{(N_{\rm BS}/N_{\rm RF},1)}_k\\
\boldsymbol{r}^{(1,2)}_k & \boldsymbol{r}^{(2,2)}_k & \cdots & \boldsymbol{r}^{(N_{\rm BS}/N_{\rm RF},2)}_k\\
\vdots & \vdots &  \ddots & \vdots\\
\boldsymbol{r}^{(1,N_{\rm UE})}_k & \boldsymbol{r}^{(2,N_{\rm UE})}_k & \cdots & \boldsymbol{r}^{(N_{\rm BS}/N_{\rm RF},N_{\rm UE})}_k
\end{matrix}
\right].
\end{equation}
Considering the uplink and downlink channel reciprocity, the absolute value of each entry of $\boldsymbol{R}_k$ is essentially the equivalent channel gain.

To maximize the equivalent channel gain in (\ref{DetermineFirst}), we just need to find the entry with the largest absolute value in $\boldsymbol{R}_k$. Denote the row index and column index of the corresponding entry as $p_k$ and $q_k$, respectively, which means that the best analog beamforming vector is $\boldsymbol{w}_c(p_k)$ and the best combining vector at the BS is $\boldsymbol{f}_c(q_k)$ for uplink transmission of the $k$th user.

According to channel reciprocity, for the downlink transmission of the $k$th user, the best analog beamforming vector at the BS is $\widetilde{\boldsymbol{f}}_k^{\rm RF} = (\boldsymbol{f}_c(q_k))^*$ and the best combining vector at the user is $\widetilde{\boldsymbol{w}}_k=(\boldsymbol{w}_c(p_k))^*$. Then the designed analog precoder at the BS is
\begin{equation}\label{AnalogPrecoding}
\widetilde{\boldsymbol{F}}_{\rm RF}=[\widetilde{\boldsymbol{f}}_1^{\rm RF}, \widetilde{\boldsymbol{f}}_2^{\rm RF}, \ldots,\widetilde{\boldsymbol{f}}_K^{\rm RF}].
\end{equation}

Compared to the beam training scheme based on exhaustive search, in the OP-based scheme, all the users can perform beam training simultaneously with the BS. This parallel manner can reduce the number of beam training times from $N_{\rm BS } N_{\rm UE}K / N_{\rm RF}$ to $N_{\rm BS} N_{\rm UE} / N_{\rm RF}$, e.g., from 512 to 64 for $N_{\rm BS }=64$, $N_{\rm UE}=8$ and $N_{\rm RF}=8$.

Note that although the beam training scheme based on hierarchial codebook~\cite{xiao2017channel} can reduce the overhead for a single user, it cannot be performed in parallel between the BS and multiple users. Moreover, there is an overhead caused by transmission of additional pilot signal for channel estimation even after the beam training is finished.

\subsection{Channel Estimation and Digital Precoder Design}\label{subsec.ChannelEstimation}
No extra pilot signal is required for channel estimation. It is based on the result from the beam training stage.

Denote
\begin{equation}\label{effectivechannel}
 {\boldsymbol{\bar{H}}}=\left[
\begin{matrix}
\boldsymbol{w}^H_1\boldsymbol{H}^{\rm dl}_1\boldsymbol{f}^{\rm RF}_1 & \boldsymbol{w}^H_1\boldsymbol{H}^{\rm dl}_1\boldsymbol{f}^{\rm RF}_2 & \cdots & \boldsymbol{w}^H_1\boldsymbol{H}^{\rm dl}_1\boldsymbol{f}^{\rm RF}_K\\
\boldsymbol{w}^H_2\boldsymbol{H}^{\rm dl}_2\boldsymbol{f}^{\rm RF}_1 & \boldsymbol{w}^H_2\boldsymbol{H}^{\rm dl}_2\boldsymbol{f}^{\rm RF}_2 & \cdots & \boldsymbol{w}^H_2\boldsymbol{H}^{\rm dl}_2\boldsymbol{f}^{\rm RF}_K\\
\vdots & \vdots &  \ddots & \vdots\\
\boldsymbol{w}^H_K\boldsymbol{H}^{\rm dl}_K\boldsymbol{f}^{\rm RF}_1 & \boldsymbol{w}^H_K\boldsymbol{H}^{\rm dl}_K\boldsymbol{f}^{\rm RF}_2 & \cdots & \boldsymbol{w}^H_K\boldsymbol{H}^{\rm dl}_K\boldsymbol{f}^{\rm RF}_K
\end{matrix}
\right].
\end{equation}
According to (\ref{finaldata2}), we have
\begin{align}\label{finaldata3}
\hat{s}_{k} =[\boldsymbol{\bar{H}}]_{k,:} \boldsymbol{F}_{\rm BB}\boldsymbol{s}+\boldsymbol{w}^{H}_{k}\boldsymbol{n}^{\rm dl}_{k},~k=1,2,\ldots,K.
\end{align}
Denote $\hat{\boldsymbol{s}} \triangleq [\hat{s}_{1},\hat{s}_{2},\ldots,\hat{s}_{K}]^T $, then
\begin{equation}\label{finaldata4}
\hat{\boldsymbol{s}} = \boldsymbol{\bar{H}} \boldsymbol{F}_{\rm BB}\boldsymbol{s} + \boldsymbol{n}^{\rm dl},
\end{equation}
where $\boldsymbol{n}^{\rm dl} \triangleq [\boldsymbol{w}_1^H \boldsymbol{n}^{\rm dl}_1 ,\boldsymbol{w}_2^H \boldsymbol{n}^{\rm dl}_2,\ldots,\boldsymbol{w}_K^H \boldsymbol{n}^{\rm dl}_K ]^T$. From (\ref{finaldata4}), the design of $\boldsymbol{F}_{\rm BB}$ relies on the estimation of $\boldsymbol{\bar{H}}$.

In fact, we can estimate $\boldsymbol{\bar{H}}$ based on $\boldsymbol{R}_1$, $\boldsymbol{R}_2$, ..., $\boldsymbol{R}_K$ in (\ref{AllProcessedPilotSignal}). Denote the estimate of $\boldsymbol{\bar{H}}$ as ${\widetilde{\boldsymbol{H}}}$. The entry on the $i$th row and $j$th column of ${\widetilde{\boldsymbol{H}}}$ can be expressed as
\begin{equation}\label{ChannelEstimation}
  [{\widetilde{\boldsymbol{H}}}]_{i,j}=[\boldsymbol{R}_i]_{p_i,q_j}
\end{equation}
where $p_i$ and $q_j$ have already been determined during the beam training.


Once channel is estimated, the ZF digital precoder and MMSE digital precoder will be
\begin{equation}\label{ZFChannelEstimate}
\boldsymbol{F}^{\rm ZF}_{\rm BB}={\widetilde{\boldsymbol{H}}}^H({\widetilde{\boldsymbol{H}}}{\widetilde{\boldsymbol{H}}}^H)^{-1},
\end{equation}
and
\begin{equation}\label{MMSEChannelEstimate}
\boldsymbol{F}^{\rm MMSE}_{\rm BB}={\widetilde{\boldsymbol{H}}}^H\Big(\frac{P}{K}{\widetilde{\boldsymbol{H}}}{\widetilde{\boldsymbol{H}}}^H +\sigma^2_{\rm dl}\boldsymbol{I}_K \Big)^{-1},
\end{equation}
respectively. In order to satisfy the total power constraint, each column of the designed digital precoder via (\ref{ZFChannelEstimate}) or (\ref{MMSEChannelEstimate}), denoted as $\boldsymbol{\bar{f}}^{\rm BB}_k$ should be normalized, i.e., $\widetilde{\boldsymbol{f}}^{\rm BB}_k=\boldsymbol{\bar{f}}^{\rm BB}_k / \| \widetilde{\boldsymbol{F}}_{\rm RF}\boldsymbol{\bar{f}}^{\rm BB}_k\|_F$ such that $\|\widetilde{\boldsymbol{f}}^{\rm BB}_k\|_F=1,~k=1,2,...,K$.

Now we have designed the analog precoder at the BS, the digital precoder at the BS and the analog combiner at the users as $\widetilde{\boldsymbol{F}}_{\rm RF}$, $\{\widetilde{\boldsymbol{f}}_k^{\rm BB} \}^K_{k=1}$, and $\{\widetilde{\boldsymbol{w}}_k\}^K_{k=1}$, respectively.

\section{Multiuser Beam Allocation}\label{sec.BeamAllocation}
In this section, we first introduce the beam conflict in multiuser mmWave massive MIMO communication systems. Then we propose a beam allocation algorithm, which can effectively eliminate the beam conflict among different users and improve the system performance.
\subsection{Beam Conflict}\label{subsec.BeamConflict}

\begin{figure}[!t]
\centering
\includegraphics[height=40mm,width=65mm]{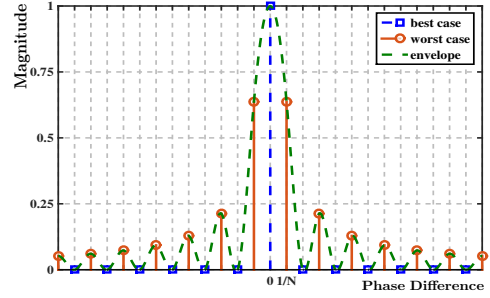}
\caption{Illustration of channel power leakage.}
\label{fig:leakage}
\end{figure}

The analog beamforming vectors considered in this work are essentially $N_{\rm BS}$ equally spaced channel steering vectors pointing at $N_{\rm BS}$ different directions. However, since the randomly distributed users may not lie in the exact directions of the $N_{\rm BS}$ beams, it generally suffers from channel power leakage. The correlation between the analog beamforming vectors and the channel steering vector at the BS can be denoted as
\begin{align}\label{InnerProduct}
C(n)&\triangleq \big| \boldsymbol{a}^H_{\rm BS}(\phi)\boldsymbol{f}^*_c(n) \big| \notag\\
&= \big| \boldsymbol{u}^H(N_{\rm BS},\phi)\boldsymbol{u}^*(N_{\rm BS},-1+(2n-1)/N_{\rm BS}) \big| \notag\\
&=\frac{1}{N_{\rm BS}}\Bigg|\frac{{\rm sin}\frac{\pi {N_{\rm BS}} \big[\phi+(-1+\frac{(2n-1)}{N_{\rm BS}})\big]}{2}}{{\rm sin}\frac{\pi\big[\phi+(-1+\frac{(2n-1)}{N_{\rm BS}})\big]}{2}}\Bigg|,n=1,2,...,N_{\rm BS}.
\end{align}
which is also illustrated in Fig.~\ref{fig:leakage}. If the analog precoding vector happens to point at the channel steering vector, i.e., $\phi \in \{-1+\frac{(2m-1)}{N_{\rm BS}},m=1,2,...,N_{\rm BS}\}$, the envelop of $C(n)$ will appear as only a single peak in the main lobe, which corresponds to the case without any channel leakage. Otherwise, it will appear as two high peaks around the main lobe as well as several low peaks in the side lobes. In the worst case, i.e., $\phi\in \{-1+\frac{2m}{N_{\rm BS}},m=1,2,...,(N_{\rm BS}-1)\}$, the envelop of $C(n)$ will appear as two equally high peaks in the main lobe as well as several high peaks in the side lobes.

According to \eqref{BSbeamformingvector}, the BS codewords in $\boldsymbol{\mathcal{F}}_c$ divide the signal coverage of the BS into $N_{\rm BS}$ sectors, where each sector is denoted as $\boldsymbol{S}_m=[-1+(2m-2)/N_{\rm BS}, -1+2m/N_{\rm BS}],m=1,2,...,N_{\rm BS}$. We denote the AoD of the dominant channel path between the BS and the $k$th user as $\phi_k$. If $\phi_k$ is in the sector $\boldsymbol{S}_{\widetilde{m}}$, i.e., $\phi_k \in [-1+(2\widetilde{m}-2) / N_{\rm BS},-1+2\widetilde{m}/N_{\rm BS}]$, the BS codeword $\boldsymbol{f}_c(\tilde{m})$ will be selected to form an analog beam serving the $k$th user. Once two users lie in the same sector, such as they are geographically close to each other, the same BS codeword forming the same beam will serve two different users, leading to the beam conflict.

As shown in \eqref{effectivechannel}, if two users are served by the same BS codeword, there will be two same columns in $\boldsymbol{\bar{H}}$, which makes it low-rank. In this case, no matter how we design the digital precoder $\boldsymbol{F}_{\rm BB}$ at the BS, the product between $\boldsymbol{F}_{\rm BB}$ and $\boldsymbol{\bar{H}}$ is low-rank and thus can not be diagonalized, which causes severe interference among different users and reduces the sum-rate.

Note that $\phi_k$ obeys the uniform distribution in $[-1,1]$ in general. Therefore, the probability of each  BS codeword to be selected to serve the $k$th user is $1/N_{\rm BS}$. The probability that all the $K$ users are served by different BS codewords without any beam conflict is equivalent to the probability of selecting $K$ different BS codewords from total $N_{\rm BS}$ codewords, which equals
\begin{equation}\label{ProbabilityBeamConflict}
   \mathcal{P}_{\rm NC}= \frac{N_{\rm BS}!}{N_{\rm BS}^K(N_{\rm BS}-K)!}.
\end{equation}
Therefore, the probability that there exists beam conflicts is $\mathcal{P}_{\rm C} =1-\mathcal{P}_{\rm NC}$. For a typical mmWave massive MIMO system with $N_{\rm BS}=64$ and $K=10$, we have $\mathcal{P}_{\rm C}\approx 52.3\%$. If we further increase $K$ to $16$, we have $\mathcal{P}_{\rm C}\approx 87.1\%$. Therefore, as the number of users increases, the same beam from the BS may point at different users, leading to higher probability of beam conflict and more severe multiuser interference.

\subsection{Beam Allocation}\label{subsec.BeamAllocationAlgorithm}
To eliminate the beam conflict and therefore improve the sum-rate, we consider beam allocation for different users. In the following, we will show that each user has several alternative beams that can be used for the beam allocation even if the channel has only one path.

\begin{figure}[!t]
\centering
\includegraphics[width=65mm,height=42mm]{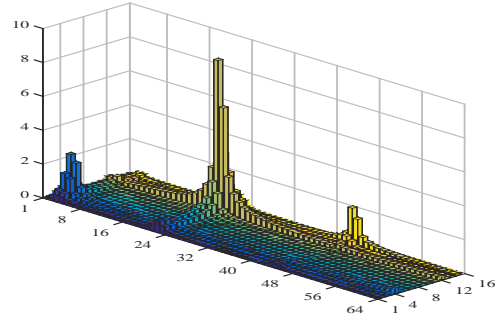}
\caption{Illustration of virtual channel gain with $N_{\rm BS}=64$ and $N_{\rm UE}=16$.}
\label{fig:EffectiveChannelGains}
\end{figure}

We define $\boldsymbol{H}^v_k \in \mathbb{C}^{N_{\rm UE}\times N_{\rm BS}}$ as the virtual channel matrix,
\begin{equation}\label{VirtualChannelMatrix}
\boldsymbol{H}^{v}_k=\boldsymbol{\mathcal{W}}^T_c\boldsymbol{H}^{\rm dl}_k\boldsymbol{\mathcal{F}}^*_c.
\end{equation}
The absolute value of the entry on the $i$th row and the $j$th column of $\boldsymbol{H}^v_k$ represents the virtual channel gain for noiseless uplink transmission of the $k$th user when using $\boldsymbol{w}_c(i)$ and $\boldsymbol{f}_c(j)$ for analog beamforming at the user and analog combining at the BS, respectively. In fact, $\boldsymbol{R}_k$ equals $\boldsymbol{H}^v_k$ if there is no channel noise. Therefore, maximize equivalent channel gain in (\ref{DetermineFirst}) is essentially maximize the virtual channel gain.

Fig.~\ref{fig:EffectiveChannelGains} illustrates the virtual channel gain for an example with $N_{\rm BS}=64$ and $N_{\rm UE}=16$, where the columns and the rows of virtual channel matrix are represented by the $x$ axis and $y$ axis, respectively. From the figure, the gain of most entries is smaller than 1. Only one or two entries are larger than 4. The relatively small entries besides the peak entry indicate the effect of channel power leakage. In fact, the power of each channel path is mainly concentrated on the intersection of two adjacent rows and two adjacent columns of the virtual channel matrix~\cite{ma2017channel}. Therefore, there are several alternative beams with large virtual channel gain even if the channel has only one path. This inspires us to eliminate beam conflict through proper beam allocation.


Considering beam conflict, (\ref{DetermineFirst}) can be reformulated as
\begin{align}\label{BeamAllocationProblem1}
& \max_{\{\boldsymbol{f}^{\rm RF}_k\}_{\rm k=1}^K, \{\boldsymbol{w}_k\}_{\rm k=1}^K}\Big\{ |\boldsymbol{w}^H_k\boldsymbol{H}^{\rm dl}_k\boldsymbol{f}_k^{\rm RF}|\Big\}_{\rm k=1}^K,\\
&~~{\rm s.t.}~~~\boldsymbol{w}_k\in\boldsymbol{\mathcal{W}}_c,~\boldsymbol{f}^{\rm RF}_k\in\boldsymbol{\mathcal{F}}_c,\label{SelectedFromCodeBook1} \\
&~~~~~~~~~ \boldsymbol{f}^{\rm RF}_i\neq \boldsymbol{f}^{\rm RF}_j, i,j=1,2,...,K,~ i\neq j. \label{NonUserConflict1}
\end{align}
The objective function in (\ref{BeamAllocationProblem1}) is the same as that of (\ref{DetermineFirst}). Note that (\ref{DetermineFirst}) is essentially $K$ independent optimization problems. However, according to the constraint in (\ref{NonUserConflict1}), the beams allocated for different users should be also different, implying that there is no beam conflict for different users. Therefore, (\ref{NonUserConflict1}) introduces inner relations among $K$ optimization problems and converts it into a multi-objective optimization problem~\cite{marler2004survey}. As a result, a set of Pareto optimal solutions instead of a single one are usually obtained. Therefore, optimization preference is needed to determine a proper solution from a set of solutions. Generally, we maximize the number of simultaneously served users by the BS, under the premise that these users satisfy the QoS. Therefore, we set the optimization preference as
\begin{equation}\label{OptimizationPreference}
\max_{\{\boldsymbol{\widetilde{f}}^{\rm RF}_k\}_{\rm k=1}^K, \{\boldsymbol{\widetilde{w}}_k\}_{\rm k=1}^K} \sum_{k=1}^{K} \mathcal{I}\big(|\boldsymbol{\widetilde{w}}^H_k\boldsymbol{H}^{\rm dl}_k\boldsymbol{\widetilde{f}}_k^{\rm RF}| , \gamma_k \big),
\end{equation}
where
\begin{equation}\label{ThresholdFunction}
  \mathcal{I}(x, y )=u(x-y)
\end{equation}
is a binary decision function and $u(n)$ is a unit step function, $\gamma_k$ is a threshold related to the quality-of-service (QoS) for the $k$th user. In practice, different users may have different QoS constraints. For example, an user demanding live video service is constrained by a large $\gamma_k$ while an user demanding audio service is only constrained by a small $\gamma_k$.  In this optimization preference, we require that the number of users satisfying QoS constraints, i.e., whose equivalent channel gains are greater than $\gamma_k$, is maximized. With this optimization preference, the beam allocation problem can be expressed as a multi-objective bilevel optimization problem~\cite{Sinha2018Bilevel}
\begin{align}\label{BeamAllocationProblem}
&\max_{\{\boldsymbol{f}^{\rm RF}_k\}_{\rm k=1}^K, \{\boldsymbol{w}_k\}_{\rm k=1}^K}\Big\{ |\boldsymbol{w}^H_k\boldsymbol{H}^{\rm dl}_k\boldsymbol{f}_k^{\rm RF}|\Big\}_{\rm k=1}^K  \\
&{\rm s.t.}\big\{\{\boldsymbol{f}^{\rm RF}_k\}_{\rm k=1}^K, \{\boldsymbol{w}_k\}_{\rm k=1}^K \big\}\in \notag\\
&{\rm argmax}_{\big\{\{\boldsymbol{\widetilde{f}}^{\rm RF}_k\}_{\rm k=1}^K, \{\boldsymbol{\widetilde{w}}_k\}_{\rm k=1}^K\big\}} \sum_{k=1}^{K} \mathcal{I}\big(|\boldsymbol{\widetilde{w}}^H_k\boldsymbol{H}^{\rm dl}_k\boldsymbol{\widetilde{f}}_k^{\rm RF}| , \gamma_k \big),  \label{MaxQoSNumberOfUsers}\\
&~~~~\boldsymbol{w}_k\in\boldsymbol{\mathcal{W}}_c,~\boldsymbol{f}^{\rm RF}_k\in\boldsymbol{\mathcal{F}}_c,\label{SelectedFromCodeBook2} \\
&~~~~\boldsymbol{f}^{\rm RF}_i\neq \boldsymbol{f}^{\rm RF}_j, i,j =1,2,...,K,~ i\neq j, \label{NonUserConflict2}
\end{align}
where we maximize the equivalent channel gain and the number of users satisfying the QoS in the upper level objectives (\ref{BeamAllocationProblem}) and lower level objective (\ref{MaxQoSNumberOfUsers}), respectively. Therefore, we aim at maximizing the equivalent channel gain, under the premise that the number of the QoS-satisfied users without any beam conflict is maximized. When $\gamma_1=\gamma_2=\cdots=\gamma_K=0$, (\ref{MaxQoSNumberOfUsers}) can be removed, resulting in the equivalence between the optimization problem expressed by (\ref{BeamAllocationProblem})-(\ref{NonUserConflict2}) and the optimization problem expressed by (\ref{BeamAllocationProblem1})-(\ref{NonUserConflict1}). Note that once an user's QoS cannot be satisfied, it is meaningless to continue to maximize its equivalent channel gain. Therefore, we only further maximize the equivalent channel gain for the users satisfying the QoS constraints. We should narrow the set of all users in (\ref{BeamAllocationProblem}) to a subset of those users satisfying QoS constraints. Denote
\begin{equation}\label{ThresholdFunction}
  \mathcal{T}(x, y )=xu(x-y).
\end{equation}
where $u(n)$ is a unit step function. Then the optimization problem in (\ref{BeamAllocationProblem})-(\ref{NonUserConflict2}) can be expressed as
\begin{align}\label{BeamAllocationProblemFinal}
& \max_{\{\boldsymbol{f}^{\rm RF}_k\}_{\rm k=1}^K, \{\boldsymbol{w}_k\}_{\rm k=1}^K}\Big\{\mathcal{T}\big(|\boldsymbol{w}^H_k\boldsymbol{H}^{\rm dl}_k\boldsymbol{f}_k^{\rm RF}| , \gamma_k \big)\Big\}_{\rm k=1}^K,   \\
& {\rm s.t.}\big\{\{\boldsymbol{f}^{\rm RF}_k\}_{\rm k=1}^K, \{\boldsymbol{w}_k\}_{\rm k=1}^K \big\}\in \notag\\
& {\rm argmax}_{\big\{\{\boldsymbol{\widetilde{f}}^{\rm RF}_k\}_{\rm k=1}^K, \{\boldsymbol{\widetilde{w}}_k\}_{\rm k=1}^K\big\}} \sum_{k=1}^{K} \mathcal{I}\big(|\boldsymbol{\widetilde{w}}^H_k\boldsymbol{H}^{\rm dl}_k\boldsymbol{\widetilde{f}}_k^{\rm RF}| , \gamma_k \big),  \label{MaxQoSNumberOfUsersFinal}\\
&~~~~\boldsymbol{w}_k\in\boldsymbol{\mathcal{W}}_c,~\boldsymbol{f}^{\rm RF}_k\in\boldsymbol{\mathcal{F}}_c,\label{SelectedFromCodeBookFinal} \\
&~~~~\boldsymbol{f}^{\rm RF}_i\neq \boldsymbol{f}^{\rm RF}_j, i,j =1,2,...,K,~ i\neq j. \label{NonUserConflictFinal}
\end{align}
However, the aforementioned  multi-objective bilevel optimization problem is difficult to handle.

To reduce the computational complexity on solving this problem, we suppose that the beams are sequentially allocated to different users. In this context, the user allocated beam earlier has more choices than that allocated beam later. The user will have fewer candidate beams if the priority of this user is low. Therefore, in order to maximize the number of users satisfying QoS, higher priority should be given to the user with a single candidate beam that satisfies QoS. We start the beam allocation from the user with the largest equivalent channel gain. Only when the beam conflict happens, we give the high priority to the user with a single candidate beam to maximize the number of users satisfying QoS.

Now we propose a QoS constrained (QC) beam allocation scheme, as shown in \textbf{Algorithm~1}. Note that the beams are formed by the codewords of $\boldsymbol{\mathcal{F}}_c$ and $\boldsymbol{\mathcal{W}}_c$. The beam allocation is essentially the codeword allocation. We use two vectors denoted as $\boldsymbol{b}_f$ and $\boldsymbol{u}_f$ to store the indices of the BS codewords and user codewords that we finally allocate to the BS and users, respectively. We initialize both $\boldsymbol{b}_f$ and $\boldsymbol{u}_f$ to be zero. The set of indices of users for beam allocation, denoted as $\boldsymbol{ \mathcal{K}}$, is initialized to be $\{1,2,...,K\}$. Note that the size of $\boldsymbol{ \mathcal{K}}$ gets smaller as the beams are sequentially allocated to different users.

\begin{algorithm}[!t]
	\caption{QoS Constrained Beam Allocation Algorithm}
	\label{alg1}
	\begin{algorithmic}[1]
		\STATE \textbf{Input:} $\boldsymbol{F}_c$, $\boldsymbol{W}_c$, $\{\boldsymbol{R}_k\}^K_{k=1}$.
        \STATE \textbf{Initialization}: $\boldsymbol{b}_f \leftarrow\boldsymbol{0}^{K}$, $\boldsymbol{u}_f\leftarrow\boldsymbol{0}^{K}$, $\boldsymbol{ \mathcal{K}}\leftarrow\{1,2,...,K\}$.
        \STATE Obtain $\{\boldsymbol{g}^{\rm sort}_k\}^{K}_{k=1}$, $\{\boldsymbol{b} _k\}^{K}_{k=1}$ and $\{\boldsymbol{u}_k\}^{K}_{k=1}$.
        \REPEAT
        \STATE Set $\boldsymbol{\mathcal{G}}\triangleq\{g^{\rm sort}_k(1),~k\in \boldsymbol{ \mathcal{K}}$.
        \STATE Obtain $k_{\rm max}$ via (\ref{kmax}).
        \IF{ $\|\boldsymbol{g}^{\rm sort}_{k_{\rm max}}\|_0=1$}
        \STATE $k_a \leftarrow k_{\rm max}$.
		\STATE Go to Step~16.
		\ELSIF{$\Lambda \neq \varnothing$}
        \STATE Obtain $k_c$ via (\ref{k_c}).
        \STATE $k_a \leftarrow k_c$.
        \ELSE
		\STATE $k_a \leftarrow k_{\rm max}$.
	    \ENDIF
        \STATE $b_f(k_a) \leftarrow b_{k_a}(1)$, $u_f(k_a) \leftarrow u_{k_a}(1)$.
        \STATE Update $\boldsymbol{g}^{\rm sort}_k$, $\boldsymbol{b}_k$ and $\boldsymbol{u}_k$, $k=1,2,\ldots,K$, according to (\ref{UpdateKa}),~(\ref{UpdateKabk}),~(\ref{UpdateKauk}), respectively.
        \STATE $\boldsymbol{ \mathcal{K}} \leftarrow \boldsymbol{ \mathcal{K}}  \backslash \{k_a\}$.
        \STATE Update $M_k$ via (\ref{UpdateMk}).
        \UNTIL{\{$\boldsymbol{ \mathcal{K}}=\varnothing$ or $\boldsymbol{\mathcal{G}}=\varnothing$\}}
        \STATE \textbf{Output:} $\boldsymbol{b}_f,\boldsymbol{u}_f$.
	\end{algorithmic}
\end{algorithm}

For each user, instead of only selecting the best pair $(\widetilde{\boldsymbol{f}}^{\rm RF}_k,\widetilde{\boldsymbol{w}}_k)$ that maximizes the equivalent channel gain, we select several pairs so that we have share pairs in case that the beam conflict happens. Firstly, from the $l$th column of $\boldsymbol{R}_k$, we select the entry with the largest absolute value, denoted as
\begin{equation}
  g_k(l) = \max_{i=1,2,\ldots,N_{\rm UE}} \big| [\boldsymbol{R}]_{i;l} \big|, {\rm for}\   l=1,2,\ldots,N_{\rm BS}.
\end{equation}
For each user, we find the largest equivalent channel gain corresponding to each BS codeword. We sort $\{ g_k(1),g_k(2),\ldots,g_k(N_{\rm BS}) \}$ in descending order, obtaining $\boldsymbol{g}^{\rm sort}_k$, where the largest entry of $\boldsymbol{g}^{\rm sort}_k$ is $g^{\rm sort}_k(1)$. Then we update $\boldsymbol{g}^{\rm sort}_k$ by
\begin{equation}\label{BeamGainThreshold}
  \boldsymbol{g}^{\rm sort}_k\leftarrow \Big\{g^{\rm sort}_k(i) \Big| g^{\rm sort}_k(i)\geq \gamma_k,~i \in \{1,2,\ldots,N_{\rm BS}\}\Big\}.
\end{equation}
Suppose the length of $\boldsymbol{g}^{\rm sort}_k$ is $M_k$, i.e., $M_k \leftarrow \| \boldsymbol{g}^{\rm sort}_k \|_0,~k=1,2,\ldots,K$. We denote the index of the BS codeword corresponding to $g^{\rm sort}_k(l),l=1,2,...,M_k$ in $\boldsymbol{\mathcal{F}}_c$ as $b_k(l)$, obtaining $\boldsymbol{b}_k$. We also denote the index of the user codeword corresponding to $g^{\rm sort}_k(l),l=1,2,...,M_k$ in $\boldsymbol{\mathcal{W}}_c$ as $u_k(l)$, obtaining $\boldsymbol{u}_k$. Therefore, for each BS codeword, now we find the user codeword with the largest equivalent channel gain satisfying QoS. These steps are summarized in Step~3.

Then we select the largest entry of $\boldsymbol{g}^{\rm sort}_k,~k\in \boldsymbol{ \mathcal{K}}$, forming a set $\boldsymbol{\mathcal{G}}\triangleq\{g^{\rm sort}_k(1),~k\in \boldsymbol{ \mathcal{K}}\}$. The index of the largest entry of $\boldsymbol{\mathcal{G}}$ is defined as
\begin{equation}\label{kmax}
  k_{\rm max} \triangleq \arg\max\limits_{~k\in \boldsymbol{ \mathcal{K}}}\{g^{\rm sort}_k(1)\},
\end{equation}
which corresponds to the strongest beam.

If $\|\boldsymbol{g}^{\rm sort}_{k_{\rm max}}\|_0=1$ indicating that the $k_{\rm max}$th user has only one candidate beam and we cannot allocate this beam to the other users, we set $k_a \leftarrow k_{\rm max}$ and then go to Step~16, where $k_a$ is defined as the index of the user finally allocated with this beam.


Otherwise, we check if there is beam conflict with the other users. If the conflict happens with some other users who have only one candidate beam, i.e.,
\begin{equation}\label{beamConflict}
\boldsymbol{\Lambda} \triangleq \big\{k \big|\ \|\boldsymbol{g}^{\rm sort}_{k}\|_0=1, b_k(1)=b_{k_{\rm max}}, k\in \boldsymbol{ \mathcal{K}} \backslash \{k_{\rm max}\} \big\}
\end{equation}
where $\boldsymbol{\Lambda} \neq \varnothing$, we obtain the index of the largest entry among these users as
\begin{equation}\label{k_c}
  k_c \triangleq \arg\max_{k\in \boldsymbol{\Lambda}}~g_k^{\rm sort}(1).
\end{equation}
Then we set $k_a \leftarrow k_c$. If $\boldsymbol{\Lambda} = \varnothing$ indicating there is no beam conflict with single beam users, we simply set $k_a \leftarrow k_{\rm max}$.

The indices of BS codeword and the user codeword corresponding to $g_{k_a}^{\rm sort} (1)$ are $b_{k_a}(1)$ and $u_{k_a}(1)$, respectively. Then we allocate this beam to the $k_a$th user by writing the indices of the codewords into $\boldsymbol{b}_f$ and $\boldsymbol{u}_f$, i.e., $b_f(k_a) \leftarrow b_{k_a}(1)$, $u_f(k_a) \leftarrow u_{k_a}(1)$.

Once this beam has been allocated to the $k_a$th user, we delete all the candidate beams of the $k_a$th user by setting $\boldsymbol{g}^{\rm sort}_{k_a}$, $\boldsymbol{b}_{k_a}$  and $\boldsymbol{u}_{k_a}$ empty. In addition, we have to delete this beam from the candidate beams of all the other users, as the other users can no longer be allocated with this beam. Therefore, we update $\boldsymbol{g}^{\rm sort}_k$, $\boldsymbol{b}_k$, and $\boldsymbol{u}_k$, $k\in \boldsymbol{ \mathcal{K}}$, as
\begin{align}\label{UpdateKa}
&\boldsymbol{g}^{\rm sort}_k \leftarrow \\
& \left\{
\begin{array}{ll}
\varnothing,~{\rm \textbf{if}}~k=k_a, \\
\boldsymbol{g}^{\rm sort}_k\backslash\big\{g^{\rm sort}_k(i) \big| b_k(i)=b_{k_a}(1), i\in \{1,2,\ldots,M_k\}\big\},~{\rm \textbf{else}}, \notag
\end{array}
 \right.
\end{align}
\begin{align}\label{UpdateKabk}
&\boldsymbol{b}_k\leftarrow \\
&\left\{
\begin{array}{ll}
\varnothing,~{\rm \textbf{if}}~k=k_a, \\
\boldsymbol{b}_k\backslash\big\{b_k(i) \big| b_k(i)=b_{k_a}(1), i\in \{1,2,\ldots,M_k\}\big\},~{\rm \textbf{else}}, \notag
\end{array}
 \right.
\end{align}
and
\begin{align}\label{UpdateKauk}
&\boldsymbol{u}_k\leftarrow \\
&\left\{
\begin{array}{ll}
\varnothing,~{\rm \textbf{if}}~k=k_a, \\
\boldsymbol{u}_k\backslash\big\{u_k(i) \big| b_k(i)=b_{k_a}(1), i\in \{1,2,\ldots,M_k\}\big\},~{\rm \textbf{else}}, \notag
\end{array}
 \right.
\end{align}
respectively. Since the number of the users for us to allocate beams is decreased by one, we update $\boldsymbol{ \mathcal{K}}$ by
$\boldsymbol{ \mathcal{K}} \leftarrow \boldsymbol{ \mathcal{K}} \backslash \{k_a\}$. Meanwhile, we update $M_k$ as the length of $\boldsymbol{g}^{\rm sort}_k$ by
\begin{equation}\label{UpdateMk}
  M_k \leftarrow \| \boldsymbol{g}^{\rm sort}_k \|_0,~k\in \boldsymbol{ \mathcal{K}}.
\end{equation}

We repeat the above steps until one of the following two conditions is satisfied.

1) We finish the beam allocation to all users, i.e., $\boldsymbol{ \mathcal{K}}=\varnothing$. For example, two users share three beams. Once each user is allocated with a beam, it is finished.

2) The set of candidate beams is empty, i.e., $\boldsymbol{\mathcal{G}}=\varnothing$. For example, two users share a beam. Once this beam is allocated to either one of the users, it is finished since there is no candidate beam available.

Finally, we output $\boldsymbol{b}_f$ and $\boldsymbol{u}_f$, where the $k$th user is allocated with the BS codeword $\boldsymbol{f}_c (b_f(k))$ and the user codeword $\boldsymbol{w}_c(u_f(k))$.

Note that during the beam training described in Section~\ref{subsec.BeamTraining}, we find the best analog beamforming vector $\boldsymbol{w}_c(p_k)$ and the best combining vector $\boldsymbol{f}_c(q_k)$ for uplink transmission, which does not consider the beam conflict and can now be replaced by \textbf{Algorithm~1}.

\section{Partial Beam Training}\label{sec.BeamTrainingScheme}
The uplink beam training scheme presented in Section~\ref{subsec.BeamTraining} needs $N_{\rm UE} N_{\rm BS} / N_{\rm RF}$ times of beam training. Moreover, for each user, we have to find the best beam with the largest equivalent channel gain from all of $N_{\rm UE} N_{\rm BS}$ beam pairs, which consumes large storage and signal processing resources. Note that the beam allocation only considers the multiuser beam conflict while the overhead of beam training is the same. In the following, we will propose two partial beam training schemes. Instead of testing all of $N_{\rm UE} N_{\rm BS}$ beam pairs during the beam training,  partial beam training schemes only test some of of $N_{\rm UE} N_{\rm BS}$ beam pairs, which can reduce the overhead of beam training. We first propose an IS-based beam training scheme that reduces the beam training from $N_{\rm UE} N_{\rm BS} / N_{\rm RF}$ to around $N_{\rm UE} N_{\rm BS} / (2N_{\rm RF})$. Then we propose a SP-based beam training scheme where the overhead of beam training is flexible and can be set arbitrarily.

\subsection{IS-based Beam Training}\label{subsec.DotInterlacedScanning}
Instead of testing all of $N_{\rm UE} N_{\rm BS}$ pairs of codewords during uplink beam training to figure out all the entries of $\boldsymbol{R}_k,k=1,2,\ldots,K$, now we propose an IS-based beam training scheme that only tests a bit more than $N_{\rm UE} N_{\rm BS}/2$ pairs of codewords to figure out around half entries of $\boldsymbol{R}_k,k=1,2,\ldots,K$, which can reduce the overhead of beam training by nearly half.

As shown in Fig.~4(a), we initially test $N_{\rm UE} N_{\rm BS}/2$ pairs of codewords to figure out half entries of $\boldsymbol{R}_k$ while setting the other untested entries of $\boldsymbol{R}_k$ zero, $k=1,2,\ldots,K$, i.e.,
\begin{equation}\label{DISeffectivechannelmatrix}
\big[ \boldsymbol{R}_k \big]_{i,j}=\left\{
\begin{array}{ll}
{\rm initially~tested~entry},&{\rm if}~(i+j) {\rm ~is~odd}, \\
0,&\rm{else},
\end{array}
 \right.
\end{equation}
where $\boldsymbol{R}_k,~k=1,2,\ldots,K$ is defined in (\ref{AllProcessedPilotSignal}). This step is indicated by Step~2 of \textbf{Algorithm~2}. We introduce a temporary matrix $\boldsymbol{R}_k^{\rm IS}$, which is initialized to be $\boldsymbol{R}_k,~k=1,2,\ldots,K$.

Since the channel power leakage is mainly concentrated on two adjacent channel entries for one dimensional channel as shown in Fig.~\ref{fig:leakage}, the power of each channel path is mainly concentrated on two adjacent rows and two adjacent columns of channel matrix, which can be observed from Fig.~\ref{fig:EffectiveChannelGains}. The largest entry of $\boldsymbol{R}_k$ indicating the largest virtual channel gain is included in the four entries on the intersection of the above two adjacent rows and two adjacent columns. In this scheme, two entries on the intersection can be obtained from the initial test.

\begin{figure}[!t]
\centering
\includegraphics[width=65mm,height=75mm]{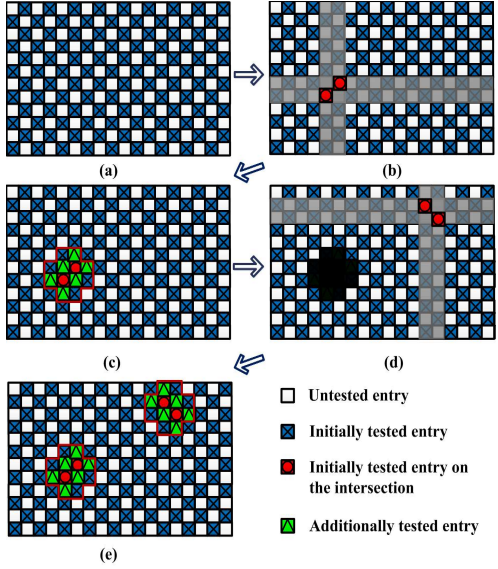}
\caption{Illustration of IS-based beam training scheme.}
\label{fig:DIS}
\end{figure}
As shown in Fig.~4(b), we find two adjacent rows indexed by $\{p_k^{\rm IS}, p_k^{\rm IS}+1\}$ with the largest average power from $\boldsymbol{R}_k^{\rm IS}$ as
\begin{equation}\label{RowCoordinates}
p_k^{\rm IS} = \arg \max_{p=1,2,...,N_{\rm UE}-1} \frac{ \big\| \big[ \boldsymbol{R}_k^{\rm IS} \big]_{p,:} \big\|_2 + \big\| \big[ \boldsymbol{R}_k^{\rm IS} \big]_{p+1,:} \big\|_2 }{\big\| \big[ \boldsymbol{R}_k^{\rm IS} \big]_{p,:} \big\|_0 +  \big\| \big[ \boldsymbol{R}_k^{\rm IS} \big]_{p+1,:} \big\|_0 },
\end{equation}
where the $\ell_2$ norm is to obtain the total power of a row while the $\ell_0$ norm is to obtain the number of the nonzero entries. Similarly, we find $\{q_k^{\rm IS},q_k^{\rm IS}+1\}$ corresponding to the columns.
The two adjacent rows and columns are marked in grey in Fig.~4(b). Then the coordinates of four entries on the intersection can be denoted as $(p_k^{\rm IS},q_k^{\rm IS})$, $(p_k^{\rm IS}+1,q_k^{\rm IS})$, $(p_k^{\rm IS},q_k^{\rm IS}+1)$ and $(p_k^{\rm IS}+1,q_k^{\rm IS}+1)$. Note that only two entries among the above four entries are initially tested, which are marked in red with a circle in the centre in Fig.~4(b). In order to find the largest entry indicating the largest virtual channel gain, two untested entries among the above four entries need to be additionally tested. In fact, due to the noise that causes errors in finding the largest entry, all of the six adjacent entries around the two red entries need to be additionally tested, which are marked in green with a triangle in the centre. As shown in Fig.~4(c), the twelve entries including the initially tested six entries and the other to-be-additionally-tested six entries, forms the shape of a small cross, where the boundary of the cross is marked in red. Note that the cross occupies most power of the corresponding channel path, e.g., around 80\% for single path channel.

To acquire more candidate beams for beam allocation that has already been addressed in Section IV, we may search the other cross corresponding to the other channel path. The number of the cross we wanted to search is denoted as $T$, which is an input to \textbf{Algorithm~2}. After finishing the search of a cross, we have to set all the twelve entries within the cross to be zero in $\boldsymbol{R}_k^{\rm IS}$, so that we will not get the same cross in the next search based on $\boldsymbol{R}_k^{\rm IS}$. As shown in Fig.~4(d), the entries set to be zero are marked in black. Then we repeat the same procedures to find another intersection and the corresponding cross, which is illustrated in Fig.~4(e).

\begin{algorithm}[!t]
	\caption{IS-based Beam Training Scheme}
	\label{alg2}
	\begin{algorithmic}[1]
		\STATE \emph{Input:} $K$, $T$.
	    \STATE Obtain $\boldsymbol{R}_k,~k=1,2,\ldots,K$ via~(\ref{DISeffectivechannelmatrix}).
        \STATE Initialize $\boldsymbol{R}^{\rm IS}_k$ to be $\boldsymbol{R}_k$, $k=1,2,\ldots,K$.
        \FOR{$t=1,2,\ldots,T$}
        \FOR{$k=1,2,\ldots,K$}
		\STATE Obtain $p^{\rm IS}_k$ and $q^{\rm IS}_k$ via (\ref{RowCoordinates}).
       \STATE Set all the entries within the cross to be zero in $\boldsymbol{R}^{\rm IS}_k$.
        \ENDFOR
        \STATE BS simultaneously transmits the row index of the agreed entry to each user.
        \STATE Additional tests are performed to compute green entries which are then stored into $\boldsymbol{R}_k,~k=1,2,\ldots,K$.
        \ENDFOR
        \STATE \emph{Output:} $\{\boldsymbol{R}_k\}^K_{k=1}$.
	\end{algorithmic}
\end{algorithm}

To inform the users which codewords should be used for the additional test, the BS needs to transmit the row indices of the six green entries to the users, which has very limited overhead. After all the users are informed, additional tests of uplink beam training using six pairs of codewords are performed to obtain the value of the six green entries. Then the obtained six entries are stored into $\boldsymbol{R}_k,~k=1,2,\ldots,K$, indicated by Step~10. Note that $\boldsymbol{R}_k$ is always getting fulfilled with increased number of nonzero entries, while the number of nonzero entries of $\boldsymbol{R}_k^{\rm IS}$ is always reducing for the new search.

Finally we output $\{\boldsymbol{R}_k\}^K_{k=1}$, which can be directly used as the input of \textbf{Algorithm~1} to make multiuser beam allocation.

\subsection{SP-based Beam Training Scheme}\label{subsec.ProbabilityTransform}
The IS-based beam training scheme can reduce the overhead of beam training from $N_{\rm UE} N_{\rm BS} / N_{\rm RF}$ to around $N_{\rm UE} N_{\rm BS} / (2N_{\rm RF})$. In the SP-based beam training scheme, the overhead of beam training is flexible.


We define a selection-probability vector to indicate the selection probabilities of user codewords in $\boldsymbol{\mathcal{W}}_c$ during the $d$th beam training as
\begin{equation}\label{SelectionProbability}
\boldsymbol{s}_d = c_d[p_d(1),p_d(2),...,p_d(N_{\rm UE})]
\end{equation}
where $c_dp_d(n),n=1,2,...,N_{\rm UE}$ denotes the selection probability of the $n$th user codeword in $\boldsymbol{\mathcal{W}}_c$, i.e., $\boldsymbol{w}_c(n)$. $c_d$ is a scalar to ensure that the sum of all entries in $\boldsymbol{s}_d$ is 1. We initialize $p_1(1)=p_1(2)=\cdots=p_1(N_{\rm UE})=1/N_{\rm UE}$ with $c_1=1$ for the first beam training.
Note that all the users employ the same codeword during the $d$th beam training while the BS distinguishes different users based on orthogonal pilot sequences via (\ref{ProcessedPilotSignal}). Furthermore, $N_{\rm RF}$ different BS codewords occupying all RF chains can be simultaneously used to compute $N_{\rm RF}$ different entries of $\boldsymbol{R}_k,~k=1,2,\ldots,K$. After these $N_{\rm RF}$ entries are tested, we have to make a record so that the future tests will be made on the other entries of $\boldsymbol{R}_k$ instead of on these $N_{\rm RF}$ entries again. Define a matrix
\begin{equation}\label{SelectableBSCodewords}
\boldsymbol{Z}_d=[\boldsymbol{z}_d(1),\boldsymbol{z}_d(2),\ldots,\boldsymbol{z}_d(N_{\rm UE})]
\end{equation}
to record the BS codewords for the untested entries of $\boldsymbol{R}_k$, where $\boldsymbol{z}_d(n)$ denotes the indices of the BS codewords corresponding to the untested entries with the user codeword $\boldsymbol{w}_c(n),~n=1,2,\ldots,N_{\rm UE}$ during the $d$th beam training. We can initialize $\boldsymbol{Z}_1$ as $\boldsymbol{z}_1(1)=\boldsymbol{z}_1(2)=\cdots=\boldsymbol{z}_1(N_{\rm UE})=\{1,2,\ldots,N_{\rm BS}\}$ for the first beam training. Note that we can also make other kind of initialization for $\boldsymbol{Z}_1$, which will be addressed at the end of this subsection.

\begin{figure}[!t]
\centering
\includegraphics[width=70mm]{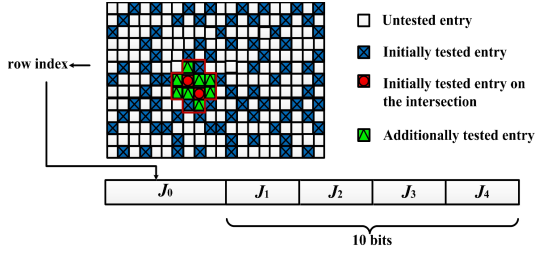}
\caption{Illustration of compressed transmit format for SP-based beam training scheme.}
\label{fig:feedback}
\end{figure}

For the $d$th beam training, based on $\boldsymbol{s}_d$ in (\ref{SelectionProbability}), we can obtain a user codeword, which is assumed to be $\boldsymbol{w}_c(n)$. Suppose we use $N_{\rm RF}$ BS codewords corresponding to $N_{\rm RF}$ indices selected from $\boldsymbol{z}_d(n)$ with equal probability, to form $N_{\rm RF}$ different transceiving codeword pairs with $\boldsymbol{w}_c(n)$ so that $N_{\rm RF}$ different entries of $\boldsymbol{R}_k$ can be tested. The set of the $N_{\rm RF}$ indices selected from $\boldsymbol{z}_d(n)$ with equal probability is denoted as $\boldsymbol{v}_d$. We can update $\boldsymbol{z}_{d+1}(n)$ as
\begin{equation}\label{UpdateBScodewords}
  \boldsymbol{z}_{d+1}(n) = \boldsymbol{z}_{d}(n) \setminus \boldsymbol{v}_d
\end{equation}
which indicates that we can no longer select the BS codewords indexed by $\boldsymbol{v}_d$ to pair with $\boldsymbol{w}_c(n)$ again. The selection probability of $\boldsymbol{w}_c(n)$, i.e., $c_d p_d(n)$ should be decreased since there are fewer BS codewords to pair with it in terms of untested entries of $\boldsymbol{R}_k$. In the extreme cases that all the BS codewords have been paired with $\boldsymbol{w}_c(n)$, the selection probability of $\boldsymbol{w}_c(n)$ is zero. If $\boldsymbol{w}_c(n)$ has been totally selected $q$ times after finishing the $d$th beam training, then there will be $N_{\rm BS}-q N_{\rm RF}$ BS codewords that have not been paired with $\boldsymbol{w}_c(n)$ in terms of untested entries of $\boldsymbol{R}_k$. Now we update the selection probability of $\boldsymbol{w}_c(n)$ for the $(d+1)$th beam training as
\begin{equation}\label{ProbabilityTransform}
p_{d+1}(n)=p_d(n)\bigg( \frac{N_{\rm BS}-q N_{\rm RF} }{N_{\rm BS}-(q-1)N_{\rm RF}}  \bigg).
\end{equation}
For the user codewords other than $\boldsymbol{w}_c(n)$, we keep their selection probability the same, i.e.,
\begin{equation}
p_{d+1}(i)=p_d(i),~i=1,2,\ldots,N_{\rm UE},~i \neq n.
\end{equation}
The selection probabilities of user codewords in $\boldsymbol{\mathcal{W}}_c$ for the $(d+1)$th beam training can be updated as
\begin{equation}\label{UpdateSelectionProbability}
\boldsymbol{s}_{d+1}=c_{d+1}[p_{d+1}(1),p_{d+1}(2),\ldots,p_{d+1}(N_{\rm UE})]
\end{equation}
where
\begin{equation}
  c_{d+1}=\frac{1}{\sum^{N_{\rm UE}}_{i=1}p_{d+1}(i)}.
\end{equation}

According to (\ref{UpdateBScodewords}), we update the record of the BS codewords for the untested entries of $\boldsymbol{R}_k$ as
\begin{equation}\label{UpdateSelectableBSCodewords}
\boldsymbol{Z}_{d+1}=[\boldsymbol{z}_{d+1}(1),\boldsymbol{z}_{d+1}(2),\ldots,\boldsymbol{z}_{d+1}(N_{\rm UE})]
\end{equation}
where
\begin{equation}
  \boldsymbol{z}_{d+1}(i) = \boldsymbol{z}_{d}(i),~i=1,2,\ldots,N_{\rm UE},~i \neq n.
\end{equation}

\begin{algorithm}[!t]
	\caption{SP-based Beam Training Scheme}
	\label{alg3}
	\begin{algorithmic}[1]
        \STATE \emph{Input:} $K,T,d_{\rm max}$.
        \STATE Obtain $\boldsymbol{R}_k,~k=1,2,\ldots,K$ from ${d_{\rm max}}$ beam training.
        \STATE Initialize $\boldsymbol{R}^{\rm SP}_k$ = $\boldsymbol{R}_k$, $k=1,2,\ldots,K$.
        \FOR{$t=1,2,\ldots,T$}
        \FOR{$k=1,2,\ldots,K$}
		\STATE Obtain $p^{\rm SP}_k$ and $q^{\rm SP}_k$ via (\ref{RowCoordinates}) by replacing $\boldsymbol{R}^{\rm IS}_k$ with $\boldsymbol{R}^{\rm SP}_k$.
       \STATE Set all the entries within the cross to be zero in $\boldsymbol{R}^{\rm SP}_k$.
        \ENDFOR
        \STATE BS simultaneously transmits the information of untested entries in compressed transmit format to each user.
        \STATE Additional tests are performed to compute green entries which are then stored into $\boldsymbol{R}_k,~k=1,2,\ldots,K$.
        \ENDFOR
        \STATE \emph{Output:} $\{\boldsymbol{R}_k\}^K_{k=1}$.
	\end{algorithmic}
\end{algorithm}
Given the number of total beam training in the initial test, e.g., $d_{\rm max}~(d_{\rm max}\leq N_{\rm BS}N_{\rm UE}/N_{\rm RF})$, we select the user codewords according to $\boldsymbol{s}_1,\boldsymbol{s}_2,\ldots,\boldsymbol{s}_{d_{\rm max}}$ for the
1st, 2nd, ... and ($d_{\rm max}$)th beam training, respectively. Note that $d_{\rm max}$ can be set arbitrarily, leading to the overhead of beam training in the initial test to be flexible. The results of total beam training are illustrated in Fig.~5, where the initially tested entries of $\boldsymbol{R}_k$ are marked in blue with a cross in the centre. Note that $\boldsymbol{s}_1,\boldsymbol{s}_2,\ldots,\boldsymbol{s}_{d_{\rm max}}$ can be generated off-line for both the BS and the users, once the methods for pseudo random generation are agreed by both sides before the beam training. The procedures to initially test the entries of $\boldsymbol{R}_k,~k=1,2,\ldots,K$ are included in Step~2 of \textbf{Algorithm~3}.

\begin{table*}[!t]
\centering
\caption{Overhead comparisons for different beam training schemes.}
\label{Table1}
\begin{tabular}{p{1.5cm}p{2.8cm}p{5.0cm}p{4.5cm}}
\toprule
  Schemes        &   Initial test &   Additional test     &     Bits to inform users\\
\midrule
OP  &  $N_{\rm BS}N_{\rm UE}/N_{\rm RF}$   &   0        &   0   \\
IS &   $N_{\rm BS}N_{\rm UE}/2N_{\rm RF}$  &   $6T$       &   $T\log_2(N_{\rm UE})$   \\
SP  &  $d_{\rm max}$  &   $12T\{1-d_{\rm max}/(N_{\rm BS}N_{\rm UE}/N_{\rm RF})\}$       &   $T\{\log_2(N_{\rm UE})+\sum^4_{i=1}J_i\}$  \\
\bottomrule
\end{tabular}
\end{table*}

In \textbf{Algorithm~3} we introduce a temporary matrix $\boldsymbol{R}_k^{\rm SP}$, which is initialized to be $\boldsymbol{R}_k,~k=1,2,\ldots,K$. The procedures of \textbf{Algorithm~3} are similar as those of \textbf{Algorithm~2}. But different from Fig.~4(c), the positions of the initially untested entries within the cross in Fig.~5 are unpredictable. Therefore, to inform the users which codewords should be used for the additional test, the BS needs to transmit the row indices of the green entries to the users. Note that the number of green entries is also unknown. Instead of straightforwardly transmitting the row indices of all the green entries to the users, now we design a compressed transmit format. As shown in Fig.~\ref{fig:feedback}, the format includes five parts. The first part $J_0$ is the row index of the first line of the cross. The other four parts indicated by $J_1$, $J_2$, $J_3$ and $J_4$ are the number of green entries in the first, second, third and fourth line of the cross, respectively.
Finally we output $\{\boldsymbol{R}_k\}^K_{k=1}$, which can be directly used as the input of \textbf{Algorithm~1} to make multiuser beam allocation.

Since the overhead of the beam training for the additional test is much lower than that of the initial test, the total overhead of the SP-based scheme is mainly determined by the initial test and therefore is flexible in terms of $d_{\rm max}$. In fact, the SP-based scheme can be regarded as an extension of the IS-based scheme. If we initialize $\boldsymbol{Z}_1$ as
\begin{equation}\label{CandidateSets}
\boldsymbol{z}_1(i)=\left\{
\begin{array}{ll}
\{1,3,5,\ldots,N_{\rm BS}-1\},&{\rm if}~i {\rm ~is~odd}, \\
\{2,4,6,\ldots,N_{\rm BS}\},&\rm{else},
\end{array}
 \right.
\end{equation}
and
\begin{equation}
d_{\rm max} = \frac{ N_{\rm BS} N_{\rm UE} }{2 N_{\rm RF}},
\end{equation}
with (\ref{ProbabilityTransform}) correspondingly revised as
\begin{equation}
p_{d+1}(n)=p_d(n)\bigg( \frac{N_{\rm BS}/2-q N_{\rm RF} }{N_{\rm BS}/2-(q-1)N_{\rm RF}}  \bigg),
\end{equation}
then the SP-based scheme is equivalent as the IS-based scheme. Moreover, the candidate sets in (\ref{CandidateSets}) can be generalized to any sets. The initialization of the probability $p_1(i),~i=1,2,\ldots,N_{\rm UE}$ can also be set different if there is prior knowledge of the users, e.g., geographic information.

\subsection{Overhead Comparisons}\label{subsec.Comparisons}
As shown in Table~\ref{Table1}, we compare the overhead of three beam training schemes, including the OP-based, IS-based and SP-based schemes. In terms of the number of beam training in the initial test, the overhead of IS-based scheme is only half of that of the OP-based scheme while the overhead of SP-based scheme is flexible, i.e., $d_{\rm max}=1,2,\ldots,N_{\rm BS}N_{\rm UE}/N_{\rm RF}$. The OP-based scheme does not need any additional test since all $N_{\rm BS}N_{\rm UE}/N_{\rm RF}$ beam pairs are initially tested. The IS-based scheme needs $6T$ beam training during additional test for each user. For the SP-based scheme, the probability that one entry of $\boldsymbol{R}_k$ is initially tested is $d_{\rm max}N_{\rm RF}/(N_{\rm BS} N_{\rm UE} )$, indicating the probability that one entry is not initially tested is $1- d_{\rm max}N_{\rm RF}/(N_{\rm BS} N_{\rm UE} )$, which results in $12(1- d_{\rm max}N_{\rm RF}/(N_{\rm BS} N_{\rm UE} ))$ entries within each cross needing additional test. In order to inform the users which codewords should be used for the additional test, the BS needs to transmit $T\log_2(N_{\rm UE})$ and $T(\log_2(N_{\rm UE})+\sum^4_{i=1}J_i)$ bits to inform each user for the IS-based and SP-based scheme, respectively. Since it is in unit of bit, meaning that the overhead of the transmission to inform users is very limited, the overhead of three schemes is mainly determined by the number of beam training in the initial test and additional test.

\section{Simulation Results}\label{sec.SimulationResults}
Now we evaluate the performance of the proposed beam training and beam allocation schemes. Consider an mmWave massive MIMO system, the number of resolvable multipath in mmWave channel is randomly set to be 3, 4 or 5 for each user, i.e., $L_k=3\sim5$, while the complex channel gain is set as $\alpha_1^k \sim \mathcal{CN}(0,1)$ and $\alpha_i^k \sim \mathcal{CN}(0,0.1)$ for $i \neq 1$. The
number of the cross we wanted to search for the IS-based and SP-based schemes is set to be $T=2$. Monte Carlo simulations are performed based on 2000 random channel implementations.

\begin{figure}[!t]
\centering
\includegraphics[width=75mm]{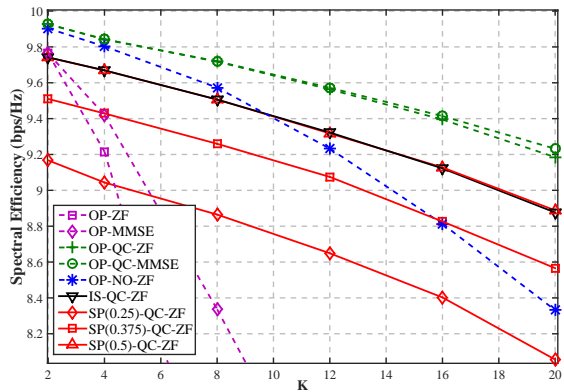}
\caption{Comparisons of spectral efficiency for different beam training and beam allocation schemes in terms of $K$.}
\label{fig:K}
\end{figure}

\begin{figure}[!t]
\centering
\includegraphics[width=75mm]{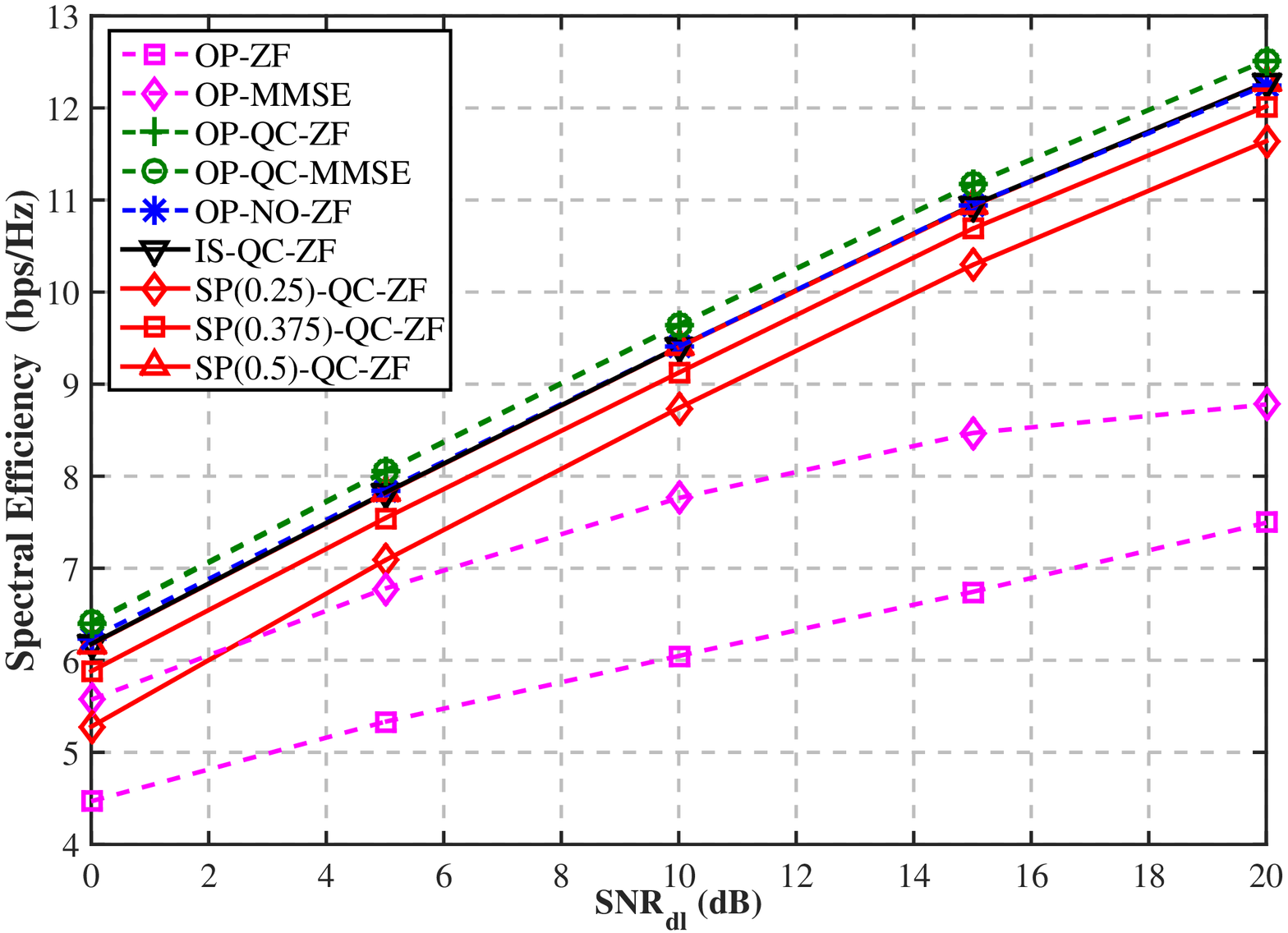}
\caption{Comparisons of spectral efficiency for different beam training and beam allocation schemes in terms of $\rm{\rm SNR_{dl}}$.}
\label{fig:SNR}
\end{figure}

\begin{figure}[!t]
\centering
\includegraphics[width=75mm]{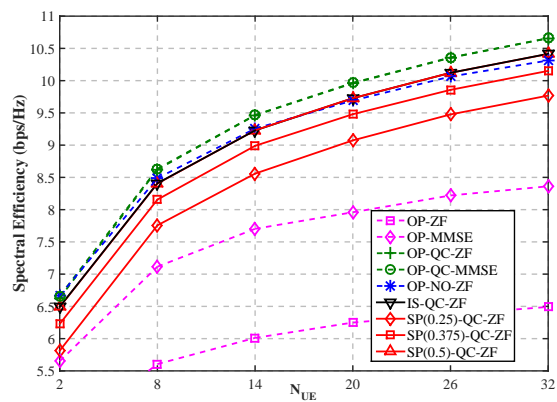}
\caption{Comparisons of spectral efficiency for different beam training and beam allocation schemes in terms of $N_{\rm UE}$.}
\label{fig:UE}
\end{figure}

\begin{figure}[!t]
\centering
\includegraphics[width=75mm]{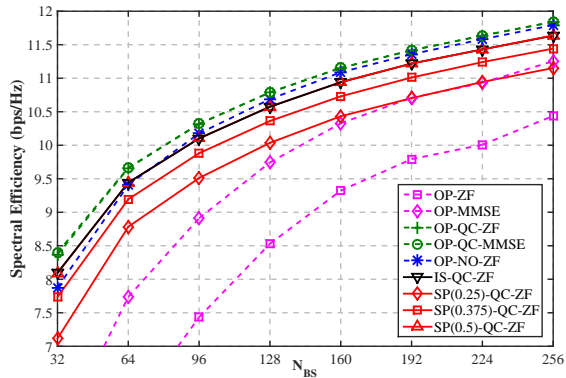}
\caption{Comparisons of spectral efficiency for different beam training and beam allocation schemes in terms of $N_{\rm BS}$.}
\label{fig:BS}
\end{figure}

The spectral efficiency illustrated from Fig.~\ref{fig:K} to Fig.~\ref{fig:BS} is defined as the sum-rate averaged over $K$. We fix the uplink channel SNR as ${\rm SNR}_{\rm ul}=10\log_{10}(\bar\alpha P_{\rm ul}/\sigma^2_{\rm ul})={\rm 20\ dB}$ for uplink beam training and channel estimation. The downlink SNR is defined as ${\rm SNR}_{\rm dl}=10\log_{10}\big(\bar\alpha P_{\rm dl}/(\sigma^2_{\rm dl} K)\big)$. For simplicity, we set $\gamma_1=\gamma_2=\cdots=\gamma_K=10\sigma_{\rm dl}$.

As shown in Fig.~\ref{fig:K}, we compare spectral efficiency for different beam training and beam allocation schemes in terms of $K$. Set $N_{\rm BS}=64$, $N_{\rm RF}=20$, $N_{\rm UE}=16$ and ${\rm SNR}_{\rm dl}={\rm 10\ dB}$. The curve labeled ``OP-ZF'' illustrates the results of the OP-based beam training scheme with analog precoding and ZF digital precoding as in (\ref{ZFChannelEstimate}). The curve labeled ``OP-MMSE'' illustrates the simulation results of the OP-based beam training scheme with analog precoding and MMSE digital precoding as in (\ref{MMSEChannelEstimate}). Note that the above two curves do not use the beam allocation to solve the problem of beam conflicts, which makes ${\boldsymbol{\bar{H}}}$ in (\ref{effectivechannel}) low rank and causes the curves to drop rapidly as $K$ increases. Since the MMSE digital precoding can slightly relief the low rank of ${\boldsymbol{\bar{H}}}$, it performs better than the ZF digital precoding. Compared to the curves of ``OP-ZF'' and ``OP-MMSE'', the curves of ``OP-QC-ZF'' and ``OP-QC-MMSE'' use the proposed QoS constrained beam allocation scheme in \textbf{Algorithm~1}, respectively. As $K$ increases, the beam conflict happens with higher probability. Once the beam conflict happens, the candidate beam with smaller equivalent channel gain is selected for one of the conflicted users, which can effectively mitigate the interference caused by the beam conflict and therefore stop the curves from fast decreasing like ``OP-MMSE'' and ``OP-ZF''. When $K=8$, the improvement of spectral efficiency of ``OP-QC-ZF'' over ``OP-ZF'' is 36.48\%, which verifies the effectiveness of beam allocation. It is observed that the curves of ``OP-QC-ZF'' and ``OP-QC-MMSE'' are almost overlapped. Therefore, once the beam conflict is treated by the beam allocation, the simple ZF digital precoding can be employed. For comparison, we also extend the near-optimal (NO) beam selection scheme proposed in \cite{gao2016near}, which is labeled as ``OP-NO-ZF''. From the figure, the proposed QC beam allocation scheme outperforms the NO scheme, e.g., 10.8\% improvement in spectral efficiency can be achieved when $K=20$. The reason is that the NO scheme selects the best beam achieving the sum-rate maximization from the group of the interference-users (IUs) at each beam selection while lacking the overall consideration for the other interference users.

\begin{table}[!t]
\centering
\caption{Simulation of overhead comparisons.}
\label{Table2}
\begin{tabular}{p{0.8cm}p{1.13cm}p{1.68cm}p{0.9cm}p{2.22cm}}
\toprule Schemes     &   Initial test &   Additional test     &     Overall &     Bits to inform users\\
\midrule
OP   &        64      &         0             &        64    &         0 \\
IS   &        32      &         12            &        44    &         8 \\
SP(0.25)  &        16      &         18            &         34    &        20 \\
SP(0.375) &        24      &         15            &         39    &        20 \\
SP(0.5)   &        32      &         12            &         44    &        20 \\
\bottomrule
\end{tabular}
\end{table}

Aside of beam allocation, we also evaluate the performance of beam training. To reduce the overhead of beam training with little sacrifice of spectral efficiency, the IS-based and SP-based schemes are introduced. The overhead comparisons of different schemes in terms of the number of beam training are provided in Table~\ref{Table2}. Since the beam allocation and digital precoding are performed by signal processing units at the BS, the overhead of the beam allocation and digital precoding are the same for different schemes. Compared to the curve of ``OP-QC-ZF'', the curve of ``IS-QC-ZF'' employs the IS-based beam training scheme that can reduce the overhead of beam training by $(64-44)/64\approx 31\%$ according to Table~\ref{Table2}, while the sacrifice of spectral efficiency is 0.29bps/Hz when $K=16$ according to Fig.~\ref{fig:K}. Since the overhead of beam training is flexible for the SP-based scheme, we initialize $\boldsymbol{Z}_1$ according to (\ref{CandidateSets}) and set $d_{\rm max}= 0.25 N_{\rm BS} N_{\rm UE} / N_{\rm RF}=16$, $d_{\rm max}= 0.375 N_{\rm BS} N_{\rm UE} / N_{\rm RF}=24$ and $d_{\rm max}= 0.5 N_{\rm BS} N_{\rm UE} / N_{\rm RF}=32$, which is the reason for the ratio in the labels ``SP(0.25)-QC-ZF'', ``SP(0.375)-QC-ZF'' and ``SP(0.5)-QC-ZF'', respectively. From the figure, as $d_{\rm max}$ increases, the spectral efficiency improves while the overhead of beam training also increases. For ``SP(0.25)-QC-ZF'', the reduction of overhead of beam training by $(64-34)/64\approx 47\%$ is achieved with the sacrifice of 1.01bps/Hz in spectral efficiency when $K=16$. For ``SP(0.375)-QC-ZF'', the reduction of overhead of beam training by $(64-39)/64\approx 39\%$ is achieved with the sacrifice of 0.59bps/Hz in spectral efficiency when $K=16$. In particular, the performance of ``SP(0.5)-QC-ZF'' is the same as that of ``IS-QC-ZF'', which verifies that the SP-based scheme is an extension of the IS-based scheme.

As shown in Fig.~\ref{fig:SNR}, we compare spectral efficiency for different beam training and beam allocation schemes in terms of $\rm{\rm SNR_{dl}}$.  We set $N_{\rm BS}=64$, $N_{\rm RF}=16$, $N_{\rm UE}=16$ and $K=10$. From the figure, the spectral efficiency of all curves increases as ${\rm SNR}_{\rm dl}$ gets larger. In particular, the performance gap between different curves keeps almost the same as ${\rm SNR}_{\rm dl}$ increases, implying that the beam allocation schemes and the beam training schemes are robust to the channel noise. Note that the multiuser interference caused by the beam conflict has already been mitigated by the beam allocation schemes, therefore the only difference comes from the channel noise.

In Fig.~\ref{fig:UE}, we compare spectral efficiency for different beam training and beam allocation schemes in terms of $N_{\rm UE}$.  We set $N_{\rm BS}=64$, $N_{\rm RF}=16$, $K=10$ and ${\rm SNR}_{\rm dl}={\rm 10dB}$. From the figure, all the curves climb up as $N_{\rm UE}$ increases, which is mainly contributed by the constant $\sqrt{N_{\rm BS} N_{\rm UE} /L_k}$ in (\ref{channelmodel}). Note that as $N_{\rm UE}$ increases, the probability of beam conflict does not change, but the gap between ``OP-NO-ZF'' and ``OP-QC-ZF'' gets larger. The reason is that as $N_{\rm UE}$ increases, the dimension of $\boldsymbol{R}_k$ in (\ref{AllProcessedPilotSignal}) gets larger, indicating there are more candidate beams with equivalent channel gain larger than the threshold for the QC scheme, while there is no benefit to the NO scheme since the NO scheme always selects the best beam with the largest equivalent channel gain.

In Fig.~\ref{fig:BS}, we compare spectral efficiency for different beam training and beam allocation schemes in terms of $N_{\rm BS}$. We set $N_{\rm UE}=16$, $N_{\rm RF}=16$, $K=10$ and ${\rm SNR}_{\rm dl}={\rm 10dB}$. Similar to Fig.~\ref{fig:UE}, all the curves climb up as $N_{\rm BS}$ increases, which is mainly contributed by the constant $\sqrt{N_{\rm BS} N_{\rm UE} /L_k}$ in (\ref{channelmodel}). From the figure, as $N_{\rm BS}$ increases, the performance gap between ``OP-NO-ZF'' and ``OP-QC-ZF'' gets smaller while ``OP-ZF'' and ``OP-MMSE'' climb faster than the others. The reason is that as $N_{\rm BS}$ grows, the number of BS codewords in $\boldsymbol{F}_c$ also increases, leading to more candidate beams and lower probability of beam conflict. Therefore, simply increasing $N_{\rm BS}$ can improve the performance, but with much higher hardware cost in practice.

\section{Conclusions}\label{sec.conclusion}
In this paper, we have proposed an OP-based beam training scheme, where all the users can simultaneously perform the beam training with the BS. We have proposed a QC-based beam allocation scheme to maximize the equivalent channel gain of the QoS-satisfied users, under the premise that the number of the QoS-satisfied users without any beam conflict is maximized. To substantially reduce the overhead of beam training, we have developed two partial beam training schemes including an IS-based scheme and a SP-based scheme. Simulation results have shown that the QC-based scheme can effectively mitigate the interference caused by the beam conflict and significantly improve the spectral efficiency while the IS-based and SP-based beam training schemes can reduce the overhead of beam training with small sacrifice of spectral efficiency. As a future work, it is of interest to explore the other performance metrics for the beam training and beam allocation schemes from some other perspectives. It is also worth developing other effective algorithms for beam allocation
as well as making the performance analysis.

\bibliographystyle{IEEEtran}
\bibliography{IEEEabrv,IEEEexample}

\end{document}